UNIVERSITY OF FUKUI
Research Center for Development of Far-Infrared Region

*FIR Center Report*

FIR UF-126　　　　　　　　　　　　　　　　October 2014

# Finite-Bandwidth Resonances of High-Order Axial Modes (HOAM) in a Gyrotron Cavity

Svilen Petrov Sabchevski and Toshitaka Idehara

*Research Center for Development of*

*Far-Infrared Region*
*University of Fukui*

*Bunkyo 3-9-1, Fukui-shi 910-8507, Japan*

# Finite-Bandwidth Resonances of High-Order Axial Modes (HOAM) in a Gyrotron Cavity

[1]Svilen Petrov Sabchevski, [2]Toshitaka Idehara

[1]*Institute of Electronics of the Bulgarian Academy of Sciences (IE-BAS),
72 Tzarigradsko Shosse Blvd., Sofia, 1784 Bulgaria,*
[2]*Research Center for Development of Far-Infrared Region,
University of Fukui (FIR FU),
Bunkyo 3-9-1, Fukui 910-8507, Japan*

**Abstract** Many novel and prospective applications of the gyrotrons as sources of coherent radiation require a broadband and continuous frequency tunability. A promising and experimentally proven technique for the latter is based on a successive excitation of a sequence of high-order axial modes (HOAM) in the cavity resonator. Therefore, the studies on HOAM are of both theoretical and practical importance. In this paper we present and discuss the methods and the results of a numerical investigation on the resonances of HOAM in a typical open gyrotron cavity. The simulations have been performed using the existing as well as novel computational modules of the problem oriented software package GYROSIM for solution of both the homogeneous and the inhomogeneous Helmholtz equation with radiation boundary conditions, which governs the field amplitude along the axis of the resonant structure. The frequency response of the cavity is studied by analyzing several resonance curves (spectral domain analysis) obtained from the numerical solution of the boundary value problem for the inhomogeneous Helmholtz equation by the finite difference method (FDM). The approach proposed here allows finite‐bandwidth resonances of HOAM to be identified and represented on the dispersion diagram of the cavity mode as bands rather than as discrete points in contrast to the frequently used physical models that neglect the finite width of these resonances. Developed numerical procedures for calculation of the field profiles for an arbitrary frequency and excitation will be embedded in the cold–cavity and self-consistent codes of the GYROSIM package in order to study the beam–wave interaction and energy transfer in gyrotron cavities.

## 1. Introduction

Discovered half a century ago [1], the gyrotron is nowadays the most powerful source of coherent CW radiation in the sub-Terahertz and Terahertz frequency bands of the electromagnetic spectrum, i.e. in a region which is still considered and referred to as a "THz power gap" or "T-gap". In recent years, the gyrotrons have demonstrated a remarkable potential for bridging this gap, situated between microwaves (electronics) and light (photonics) [2]. The progress in the development of powerful high-frequency gyrotrons has opened the way to a wide range of novel and emerging applications in the high−power THz technologies and in the physical research [3]. Some of the most significant achievements in this field have been presented recently in a special issue of the *Journal of Infrared, Millimeter and Terahertz Waves* [4] (9 papers altogether, including 3 review papers). Another class of gyrotrons that has demonstrated outstanding achievements consists of megawatt−class tubes for electron cyclotron resonance heating (ECRH) and electron cyclotron current drive (ECCD) of magnetically confined plasmas in various reactors (tokamaks and stelarators) for a controlled thermonuclear fusion [5]. Both the studies and the development of gyrotrons worldwide are based on a mature and well established theory, which provides a rich variety of adequate physical models (ranging from simplified linear and analytical to the most sophisticated self-consistent models) that describe their operation. Most of them have been implemented in numerous computer codes (e.g., MAGY [6], problem−oriented software package GYROSIM [7, 8], DAPHNE (CRPP-EPFL) [9], ARIADNE [10], ESRAY (IHM-KIT) [9, 10], CAVITY (IHM-KIT) [10], TWANG[11], EURIDICE [12], EPOS[13], just to name a few), which are used for numerical investigation, computer−aided design (CAD) and optimization of high−performance tubes. The underlying theory is well represented in several excellent monographs [14–18] as well as in an extensive literature on the subject. A driving force and a source of motivation for the further development of the simulation tools (physical models and computer codes) are the problems and challenges that are being encountered trying to satisfy the specific requirements of the novel applications. In this respect, a good example is the demand to provide a broad and continuous frequency tunability, which is essential for many advanced spectroscopic techniques (see, for example, the papers on NMR-DNP spectroscopy, XDMR, and study of the HFS of positronium in [4] as well as [19–21]).

One of the most actively discussed and successfully pursued technique for a broadband and continuous frequency tunability is based on a successive excitation of high−order axial modes (HOAM), a.k.a. longitudinal modes, in the gyrotron cavity. We have already reviewed and analyzed such possibility elsewhere [22, 23] together with other techniques for frequency control (tuning), such as frequency modulation, frequency step switching, and frequency stabilization. This concept has been explored and applied to the CAD of several gyrotrons for spectroscopic applications [22, 24]. Additionally, in [25] it has been demonstrated that a promising opportunity for widening of the tunability band is the usage of resonators in which the regular part of the cavity is replaced by a slightly up−tapered section [26]. It should be mentioned also that in most of the cases, the interaction which takes place in the cavity is the one between the electron beam and the backward−wave components of the HOAM ($TE_{m,n,-q}$), i.e. gyro–BWO [27]. Several tubes belonging to the Gyrotron FU CW series [21] also operate in broad frequency bands due to the excitation of a sequence of HOAM. For example, in the gyrotron FU CW IV a continuous tunability of the radiation frequency is performed by changing the magnetic field in a wide range from 134 GHz (at 4.9 T) to 140 GHz (at 5.2 T) operating on the $TE_{1,2,q}$ modes with axial indices $q = \pm 1, -2, -3, -4$ [28]. Another good example, which illustrates this approach, is the gyrotron FU CW VI designed for NMR-DNP spectroscopy [29]. It operates on a sequence of high-order axial modes $TE_{0,6,q}$ at the fundamental of the cyclotron frequency using a 15 T superconducting magnet. The frequency of the radiation at the lowest axial mode ($TE_{011}$) is 394.6 GHz. Frequency tunability in a 2 GHz interval is performed using a backward wave interaction and exciting modes with different axial indices $q$ by controlling the magnetic field intensity in the cavity. A recently developed tube for 600 MHz NMR-DNP spectroscopy has demonstrated a broadband continuous frequency tunability (4 GHz around 395 GHz) operating on $TE_{7,3,q}$ modes by sweeping the magnetic field [30]. For our further analysis in this paper, it is worth mentioning that not only oscillations corresponding to the integer axial indices $q = -1, -2, \ldots -10$, but also for non-integer (intermediate) values of $q$ have been observed (see Fig.2 in [30]) as well. A series of gyrotrons for NMR-DNP spectroscopy, developed at MIT have demonstrated continuous tunability in a wide frequency range utilizing the same method [31–35]. For instance, in the 250 GHz gyrotron, tunability with a 3 GHz tuning bandwidth has been achieved by implementing a long (23 mm) interaction cavity in which a series of HOAM is being excited by changing either the magnetic field of the gyrotron or the cathode potential [33]. An important observation made in the cited paper is that "for long cavities, the modes with different $q$ values are closely spaced and can form a continuum when the electron beam is present" [33]. Furthermore, in most of these studies a smooth transition between the different HOAM is reported (see, e.g., [29, 31, 33]), which, in fact, properly justifies the usage of the term "continuous frequency tunability". Detailed theoretical, numerical (using the TWANG code), and experimental studies of the gyrotrons that are being developed at CRPP-EPFL (Switzerland) as frequency tunable radiation sources for NMR-DNP spectroscopy have provided novel physical results that give additional insight on different regimes of operation (from linear to chaotic gyro-TWT and gyro-BWO) at different HOAM [36–39].

The short survey on the literature presented above, shows that the studies on HOAM are of both theoretical and practical importance for a better understanding and implementation of a broad frequency tunability of gyrotrons. The rest of the paper is organized as follows. In Sec.2 we consider a boundary value problem (BVP) for the homogeneous Helmholtz equation, which governs the amplitude of the microwave field in a typical gyrotron resonator (open cavity). The methods used for its solution and implemented in the cold cavity codes of the software package GYREOS (as well as in many other similar computer programs) are outlined briefly in Sec. 3. The numerical approach used in this study to solve the BVP for the inhomogeneous Helmholtz equation by the finite−difference method (FDM) is presented in Sec. 4. It contains also important consideration on the eigenvalue problem for the corresponding linear system of equations and on the existence of a solution of the inhomogeneous Helmholtz equation. The frequency response of an open gyrotron cavity is studied in Sec. 5, where several resonance curves have been used in order to identify the finite-bandwidth resonances of HOAM. Finally, Sec. 6 presents an outlook and the conclusions.

## 2. Helmholtz equation for the longitudinal field profile in a gyrotron cavity

In this paper, we consider a typical gyrotron cavity, which consists of a central section (the region where the electron beam interacts synchronously with the electromagnetic field) and one or more tapers at both of its ends. The interaction region is usually a part of a regular or slightly tapered hollow metallic cylinder with a circular cross section. One of the functions of the tapers is to match and couple the cavity to the rest of the system, e.g. beam tunnel at the electron gun side and the output waveguide at the opposite side. Besides this, their configuration is important for the axial field distribution and the Q factor of the entire resonant structure. Very often, for example, the taper at the entrance of the cavity is a constriction (i.e. has a narrowing cross section towards the electron gun) while at the exit of the cavity the taper is a horn like, i.e. has an increasing cross section. Such configuration allows the radiation to be coupled only in the forward direction and prevents the propagation of the wave in the opposite direction (towards the gun). Therefore, the gyrotron resonator is an *open system* in which the eigen–functions are in fact quasi–normal modes (QNMs) [40] with complex frequencies (in order to take into account the losses, i.e. energy escape) rather than normal modes of a closed cavity or a uniform waveguide.

A classical approach for an analysis of such gyrotron cavity is the one developed by Vlasov *et al.* [41]. Its underlying physical model is based on the theory of slightly irregular waveguides (i.e. waveguides with a slowly varying radius) and is also known as "a model of a slightly irregular string". The main assumption in this approach is that the structure of the transverse field (TE or TM mode) remains unchanged and only its amplitude undergoes changes along the axis of the cavity.

The axial distribution of the complex field amplitude $F_s = \mathrm{Re}(F_s) + i\,\mathrm{Im}(F_s)$ of the mode labeled by the index $s$ is governed by the following homogeneous Helmholtz equation

$$\frac{d^2 F_s}{dz^2} + k_{z,s}^2(\omega_s, z) F_s = 0, \tag{1}$$

where

$$k_{z,s}^2(\omega_s, z) = k_s^2 - k_{\perp,s}^2 = \left(\frac{\omega_s}{c}\right)^2 - k_{\perp,s}^2. \tag{2}$$

Here, $k_{z,s}$ and $k_{\perp,s}$ are the axial and the transverse components of the complex wave propagation vector $k_s = \omega_s / c$; $c$ being the speed of light in vacuum. The complex eigen–frequency $\omega_s = \mathrm{Re}(\omega_s) + i\,\mathrm{Im}(\omega_s) = \omega_{r,s} + i\omega_{i,s}$ is represented by the relation $\omega_s = \omega_{r,s}(1 + \frac{1}{2Q_{D_s}} i)$, where $Q_{D_s}$ is the diffraction Q factor of the given mode.

The transverse wave-number is $k_{\perp,s} = \frac{\chi_{mn}}{R(z)}$, where $\chi_{mn}$ is the eigenvalue of the mode TE$_{mn}$, and $R(z)$ is the radius of the cavity at the axial position $z$.

The Helmholtz equation (1) follows naturally from the Maxwell's equations assuming a time harmonic ( $\sim \exp(i\omega_s t)$ ) dependence of the fields and reducing them to a wave equation (for a detailed derivation see, for example, [42]). It suffices only to mention here that this equation applies to both TE and TE modes but the coupling between them is neglected. In its general form (1) applies to the wave propagation in various physical situations that are distinguished by the specific boundary and initial conditions pertinent to each particular case. Before formulating the boundary conditions studied in this paper it is instructive to note that the differential operator of Eq. (1) can be factored in the following form

$$L = \frac{d^2}{dz^2} + k_z^2 = \left(\frac{d}{dz} - ik_z\right)\left(\frac{d}{dz} + ik_z\right) = L_- L_+ , \qquad (3)$$

and thus the Helmholtz equation can be rewritten as

$$Lf_s = L_- L_+ F_s = 0 . \qquad (4)$$

Using the operators $L_-$ and $L_+$ it is convenient to specify the following boundary conditions at both ends (the input and the exit, at $z_{in}$ and $z_{ex}$, correspondingly) of the cavity

$$L_- F_s = 0 \text{ at } z = z_{in}; \; L_+ F_s = 0 \text{ at } z = z_{ex} . \qquad (5)$$

While (1) and (4) are two–way wave equations since they describe waves propagating in both directions, the boundary conditions (5) are specified by the so called one–way wave equations (also referred to as a wave equation in a half–space) [43-45]. The general solution of the homogeneous Helmholtz equation is a standing wave, which is a linear combination (i.e. superposition) of both a forward and a backward propagating wave. Such observation is essential since in a gyrotron cavity the electron beam can interact synchronously either with the forward or with the backward component of the wave. In the former case a gyrotron or gyro-TWT operation takes place while in the latter a gyro-BWO or gyro-BWT regime is realized.

Therefore, in the problem that we study, i.e. open gyrotron cavity, Eq. (1) is supplemented by the following, so called radiation boundary conditions (a.k.a. Sommerfeld's boundary conditions)

$$\left[\frac{dF_s}{dz} - ik_{z,s}F_s\right]_{z=z_{in}} = 0; \; \left[\frac{dF_s}{dz} + ik_{z,s}F_s\right]_{z=z_{ex}} = 0 . \qquad (6)$$

Taking into account the physical meaning of the corresponding solution however, it is more appropriate to call them an *evanescent wave* BC (at $z = z_{in}$ ) and an *outgoing wave* BC (at $z = z_{ex}$ ), respectively. Below we will omit the index $s$ that labels the mode keeping in mind that the corresponding relation applies to any possible mode in the resonator.

In order to give some additional insight into the meaning of the boundary conditions (6) it is easy to derive their more general form considering an idealized case when the field profile is a sum of simple forward and a backward propagating waves, namely

$$F(z) = F_{for} \exp(-ik_z z) + F_{back} \exp(ik_z z) . \qquad (7)$$

Defining the reflection coefficients at both ends as

$$\Gamma_{in} = \Gamma(z_{in}) = \frac{F_{for} \exp(-ik_z z_{in})}{F_{back} \exp(ik_z z_{in})} \quad ; \quad \Gamma_{ex} = \Gamma(z_{ex}) = \frac{F_{back} \exp(ik_z z_{ex})}{F_{for} \exp(-ik_z z_{ex})} , \tag{8}$$

it is straightforward to obtain the following (more general) boundary conditions

$$\frac{dF(z_{in})}{dz} = ik_z \left(\frac{1-\Gamma_{in}}{1+\Gamma_{in}}\right) F(z_{in}) \quad ; \quad \frac{dF(z_{ex})}{dz} = -ik_z \left(\frac{1-\Gamma_{ex}}{1+\Gamma_{ex}}\right) F(z_{ex}). \tag{9}$$

Obviously, these boundary conditions reduce to (6) for $\Gamma_{in} = 0$ and $\Gamma_{ex} = 0$, respectively. Therefore, suffice it to note that the approximations $\Gamma_{in} = 0$ and $\Gamma_{ex} = 0$ are consistent with the assumptions in the studied physical model, namely an evanescent wave at the entrance of the cavity and an outgoing wave at its exit.

Recalling that both the frequency and the axial wave vectors are complex, from (2) we have

$$k_z^2 = \left(\frac{\omega_r}{c}\right)^2 \left[1 - \frac{1}{4Q_D^2} - \left(\frac{R_c}{R(z)}\right)^2 + \frac{1}{Q_D} i\right], \tag{10}$$

and, correspondingly

$$k_z = \sqrt{k^2 - k_\perp^2} = \sqrt{\frac{\omega^2}{c^2} - k_\perp^2} = k_{z,r} + ik_{z,i}, \tag{11}$$

where $k_{z,r}$ and $k_{z,i}$ are the real and the imaginary part, respectively. Denoting

$$\text{Re}\left(\frac{\omega^2}{c^2}\right) = k_{\perp c}^2 + \Delta^2 , \tag{12}$$

where $k_{\perp c} = \chi_{mn} / R_c$ is the transverse cut-off wave number for a cavity with a central regular section of radius $R_c$ and $k_z$ is the axial wave number, one can obtain the following useful relations

$$k_z = k_{z,r} + ik_{z,i} = \sqrt{A + iB}, \tag{13}$$

and

$$k_{z,r} = \sqrt{\frac{1}{2}\left(\sqrt{A^2 + B^2} + A\right)} , \quad k_{z,i} = \sqrt{\frac{1}{2}\left(\sqrt{A^2 + B^2} - A\right)} \tag{14}$$

Here the following notations are used

$$A = \Delta^2 + \frac{\chi_{mn}^2}{R_c^2}\left(1 - \frac{R_c^2}{R^2(z)}\right) , \quad B = \left(\Delta^2 + \frac{\chi_{mn}^2}{R_c^2}\right)\left[Q_D\left(1 - \frac{1}{4Q_d^2}\right)\right]^{-1}. \tag{15}$$

Analogously, separating the real and imaginary parts of Eq. (1) one can replace the Helmholtz equation by two coupled second order differential equations for real functions:

$$\left[\frac{d^2}{dz^2} + \text{Re}(k^2) - k_\perp^2\right]F_r = \text{Im}(k^2)F_i$$
$$\left[\frac{d^2}{dz^2} + \text{Re}(k^2) - k_\perp^2\right]F_i = -\text{Im}(k^2)F_r \qquad (16)$$

Analogously

$$\frac{d^2 F_r}{dz^2} = -CF_r + DF_i ,$$
$$\frac{d^2 F_i}{dz^2} = -DF_r - CF_i \qquad (17)$$

where $C = \text{Re}(k_z^2) = k_{z,r}^2 - k_{z,i}^2 = A$ and $D = \text{Im}(k_z^2) = 2k_{z,r}k_{z,i} = B$. The latter form is more convenient for the numerical integration of the equation (1). Therefore, the system of equations, which is integrated in several of the cold cavity codes of GYROSIM is

$$\frac{dF_r}{dz} = F_r', \quad \frac{dF_i}{dz} = F_i'$$
$$\frac{dF_r'}{dz} = -AF_r + BF_i, \quad \frac{dF_i'}{dz} = -BF_r - AF_i \qquad (18)$$

In the above physical model, the Ohmic losses in the cavity wall can be taken into account using a modified expression for $k_z^2$, namely [46]

$$k_z^2 = \frac{1}{c^2}\left\{\omega^2 - \omega_{mn_{cut}}^2\left[1 - (1+i)\frac{\delta}{R}\left(1 + \frac{m^2}{\chi_{mn}^2 - m^2}\frac{\omega^2}{\omega_{mn_{cut}}^2}\right)\right]\right\}, \qquad (19)$$

where $\delta$ is the skin depth and $\left[1 - (1+i)\frac{\delta}{R}\left(1 + \frac{m^2}{\chi_{mn}^2 - m^2}\frac{\omega^2}{\omega_{mn_{cut}}^2}\right)\right]$ is the loss factor [46].

## 3. Different approaches for solving the Helmholtz equation

In this section, first we recall shortly several well-known analytical and numerical approaches to the solution of the Helmholtz equation before presenting the technique used by us in this study. For completeness, however, it should be mentioned that in some physical models the field profile is approximated by analytical functions rather that calculated from the wave equation. For instance, in [47, 48] the axial dependence of the amplitude is represented by a Gaussian, while in [49] additionally a sinusoidal function (that corresponds to a closed cavity) is used as well.

### 3.1 Analytical solutions of the Helmholtz equation

Since (1) is analogous to the Schrodinger's equation for a particle in a potential well with a potential $k_z^2$ it can be treated in the framework of the adiabatic JWKB (after Jeffreys, Wentzel, Kramers, Brillouin) approximation [50], which is applicable if the radius of the cavity $R(z)$ (and thus the transverse wave number $\chi_{mn}/R(z)$) is changing slowly so that the following condition is fulfilled

$$\left|\frac{dk_z}{dz}\right| << \left|k_z^2\right| \quad . \tag{20}$$

The general JWKB solution then has the form [41]

$$F = \frac{A}{\sqrt{k_z}} e^{i\int k_z dz} + \frac{B}{\sqrt{k_z}} e^{-i\int k_z dz} \quad , \tag{21}$$

where $A$ and $B$ are constants of integration to be determined by the boundary conditions. It is obvious that depending on the components of the complex value $k_z$ the terms in (21) can be either decaying or growing exponentials or harmonic functions. According to the boundary conditions formulated above, at the electron gun end of the resonant structure the waveguide is well and truly cut-off. The wave is evanescent there and its amplitude decreases exponentially towards the electron gun. Therefore, here we have

$$F = (A/\sqrt{k_z})e^{i\int k_z dz}, \tag{22}$$

assuming $\text{Im}\{k_z\} < 0$. At the output end of the resonator the solution must be an outgoing wave which is travelling in the positive direction of the $z$ axis. Correspondingly,

$$F = (B/\sqrt{k_z})e^{-i\int k_z dz}, \tag{23}$$

assuming $\text{Re}\{k_z\} > 0$.

The main drawback of the JWKB approximation is that the condition (20) is not satisfied near the cut−off sections of the cavity. One way to overcome this problem is an approach, which provides a solution expressed in terms of Airy functions [51]. This method involves an approximation of $k_z^2$ by the first term of its expansion in Taylor series. Then, using an appropriate substitution for the axial coordinate ($z \to t$) the Helmholtz equation (1) is transformed to an Airy equation ($d^2F/dt^2 - tF = 0$) with a known analytical general solution. This approach has proved to be very effective for irregular waveguides [51, 52] and tapered cavities [53, 54] but is not applicable to waveguides with a constant radius since in the latter case the variable $t$ equals zero. It has also been applied to the analysis [55−57] and synthesis [58] of gyrotron cavities containing irregular segments (e.g. slowly varying waveguide sections, truncated cones, chain of tapered cavities). Other known analytical approximations to the field structure in open gyrotron resonators utilize Hermite polynomials [59]. Another, alternative approach has been proposed by Barroso et al. [60]. Using the Langer's transformation the wave equation is represented by an approximate identical equation, which has a solution in terms of Hermite functions.

### 3.2. Numerical solutions of the Helmholtz equation for the field amplitude in an open gyrotron cavity

Since the Helmholtz equation is one of the most fundamental equations of the mathematical physics it is widely used in many numerical studies on various wave phenomena in electromagnetics, acoustic and other physical disciplines. For its solution a great number of numerical methods (finite difference, FDM; finite element, FEM; fast multipole methods, FMM, just to name a few) are used as is evident from the extensive literature on the subject. Here, we confine ourselves only to the approaches specialized to the solution of a two-point boundary value problem (Eqs. (1), and (6)), for the longitudinal field profile in a gyrotron cavity. In most of the known computer codes the so–called "method of shooting" is used. It implies an iterative procedure in which the Helmholtz equation is integrated for a sequence of guess values of the complex frequency. Using one or another searching procedure, after each iteration the discrepancy between the result and the boundary condition at the right (exit) plane $z = z_{ex}$ is checked. The process continues until the boundary condition is matched (i.e. satisfied) to a specified accuracy. Various implementations of this general algorithm differ most notably in the following three points: (i) the method used for integration of the equation; (ii) the searching procedure; and (iii) the formulation of the convergence criteria of the iterative process. For example, some codes utilize Runge-Kutta integration, while others employ the Numerov method. Most frequently, the searching procedures are based on the MINUIT software package (part of the numerical library CERNlib) for a nonlinear optimization with continuous parameters or complex root finding codes based on the Muller's method (e.g. MULLER subroutine). The root-finding procedures used to calculate the complex frequency are in fact minimization algorithms and the iterative process continues until the following convergence condition is satisfied [61]

$$|R(\omega)| = \left| \frac{F'(z_{ex}) + ik_z(z_{ex})F(z_{ex})}{F'(z_{ex}) - ik_z(z_{ex})F(z_{ex})} \right|. \tag{24}$$

This numerical approach has been implemented in many computer codes developed by different research groups worldwide and used successfully for analysis and design of gyrotron cavities. While it is beyond the scope of our paper to review the numerous papers on this subject we refer the interested reader to [42, 61–67] for more details.

Our problem-oriented software package GYROSIM (GYROtron SIMulation) for modelling and computer-aided design (CAD) of gyrotrons also contains computational modules (embedded in several cavity codes) in which the same approach is realized [8]. They are used for calculation of the eigen-frequencies, diffractive quality factor and longitudinal field profile of the resonator in a cold–cavity approximation. As an illustration, below we present the calculated field profiles that correspond to the first six eigen-frequencies and, respectively, to the first six axial modes. The cavity which we use in this example (see Fig.1) as well as in the following calculations was selected arbitrary among numerous similar configurations used in the Gyrotron FU and Gyrotron FU CW series. It is an appropriate choice because of its most simple configuration consisting of a regular section and two adjacent tapers. In fact, a cavity with such dimensions has been extensively studied as one of the possible resonator designs of the large-orbit gyrotron (LOG) with a permanent magnet [68]. Obviously, the most distinguishing feature of the field profiles shown in Fig.2 is the number of peaks (maxima), which is conventionally used as a third (axial) index $q$ in the mode designation TE$_{mnq}$. However, some remarks concerning the axial index $q$ are in order here. First, in a closed cavity $q$ has a clear physical meaning since it is an integer which equals the number of field variations in the resonator. Accordingly, the axial wave number is given by $q\pi/L$, where $L$ is the length of the cavity. In an open cavity, however this length is not determined unambiguously by the geometrical dimensions but, rather, by the spatial extent of the field profile. Therefore, it is more appropriate to call this parameter *effective cavity length* ($L_{eff}$). Recently, a novel method for estimation of $L_{eff}$ has been proposed by M. Thumm [69].

This technique has been applied to the evaluation of the Fresnel parameter in cavities of megawatt class gyrotrons for fusion. Alternatively, if $L$ is fixed (e.g., to the geometrical length of the resonator) then the value of $q$, calculated from the above expression for the axial wave number is not necessarily an integer and thus should be called *effective axial index* ($q_{eff}$). Such conventions have already been used in [22, 23] as well as by other authors. In this paper, wherever possible, we prefer to use as an independent variable directly the axial wave number, which has a clear physical meaning and is implicitly present in the equations (see above) of the physical model. Recall also that it is, namely, the axial wave number $k_z$, which is necessary in order to plot the dispersion relation $\omega = \omega(k_z)$ of the mode.

In Fig.3, the first six axial modes are shown as points on the dispersion diagram of the considered cavity mode, $TE_{4,1}$. It is clear that the illustrated approach allows one to find a discrete spectrum of axial modes. Representing them by discrete points on the dispersion curve (as in the case of a closed cavity) means that we consider zero width resonances. Next, we describe an alternative approach that allows finite-bandwidth resonances in an open cavity to be identified.

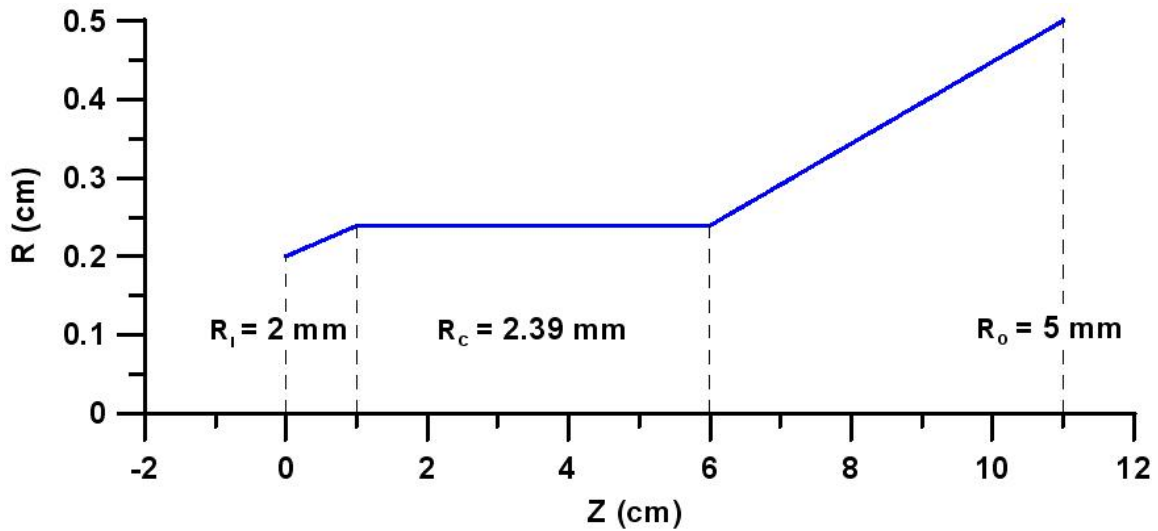

Fig. 1 Configuration and dimensions of the gyrotron cavity used in the illustrative numerical calculations.

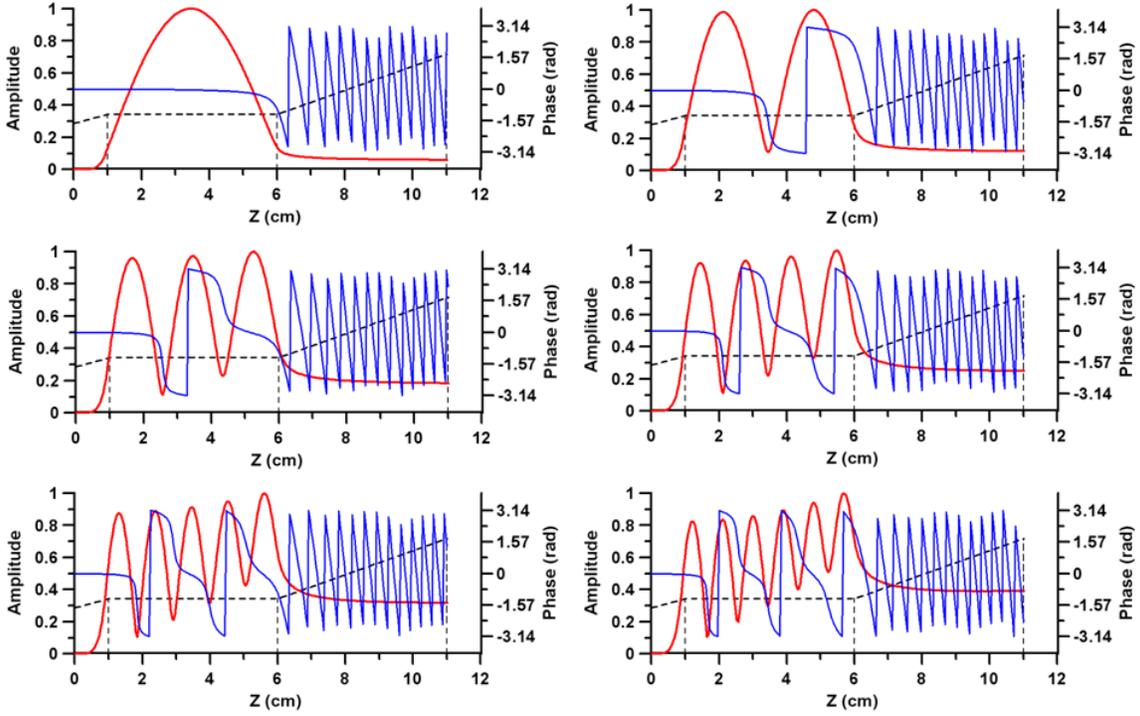

Fig. 2 Normalized amplitudes (red) and phases (blue) of the longitudinal field profiles corresponding to the first six axial modes. Dashed lines show the boundaries of the cavity sections.

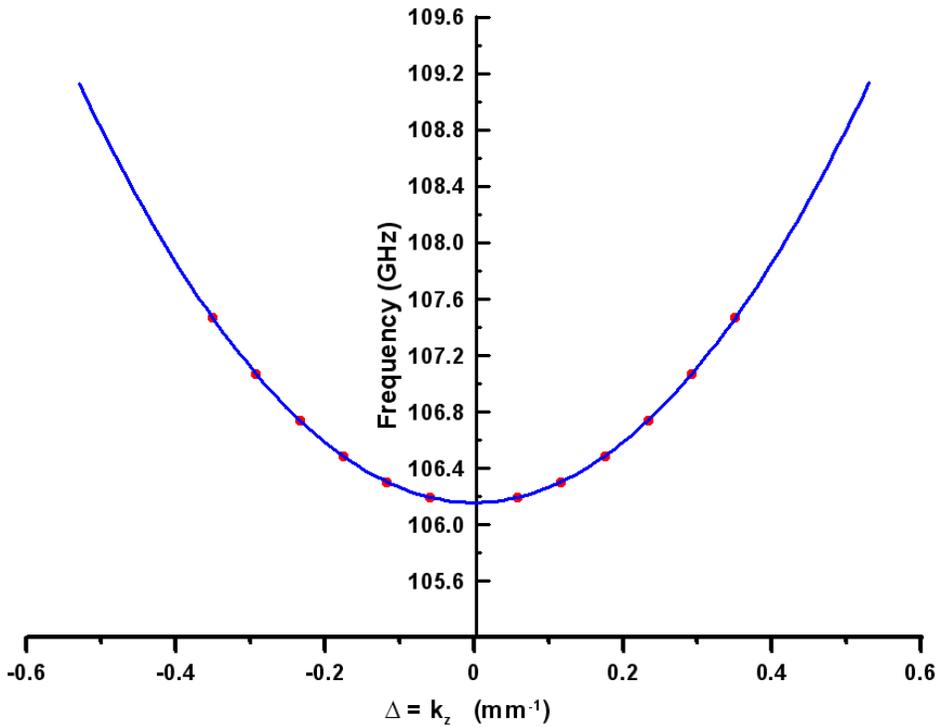

Fig. 3 First six axial resonances (red points) corresponding to the field profiles of Fig. 2 on the dispersion curve (blue line) of the mode $TE_{4,1}$.

## 4. Solution of the boundary value problem for the inhomogeneous Helmholtz equation in a gyrotron cavity by the finite differences method.

In this section we consider an inhomogeneous Helmholtz equation

$$\frac{d^2 F}{dz^2} + k_z^2(\omega, z) F = E(\omega, z), \qquad (25)$$

where the right–hand term $E(\omega, z)$ represents the excitation of the cavity. It is supplemented with the same boundary conditions as in Sec.2. While the homogeneous equation (1) is used in the cold–cavity codes its inhomogeneous counterpart (25) describes the longitudinal field profile in the self-consistent physical models, where the excitation is given by

$$E = G_b \langle P \rangle. \qquad (26)$$

Here $G_b$ is a coefficient which depends on the electron beam parameters, on the operating mode and on the used normalization of the dimensionless parameters. The complex function $\langle P \rangle$ is proportional to the energy transfer averaged over the entire ensemble of particles used in the simulations. For simplicity, the time dependence is omitted here (recall that in the time dependent cavity codes (25) is solved at each time step).

As mentioned in [14, p.59], for example, for the forced oscillations it is legitimate to use a real frequency (i.e., $\omega = \omega_r$). We adopt the same ansatz here, but as before we consider $F$ and $E$ as complex valued functions. Using a standard three point finite difference representation of the second derivative of the equation and a ghost point method [70] for the first derivatives of the boundary conditions on a grid ($z_i = i\Delta z, i = 0,1,2,...n-1$) with an axial step $\Delta z$ the BVP is reduced to a system of $n$ linear equations for $n$ unknown values of the complex field amplitudes $F_i$ at the grid points. In a matrix notation this system has the form

$$\mathbf{AF} = \mathbf{B}, \qquad (27)$$

where $\mathbf{A}$ is a tridiagonal $n \times n$ matrix, $\mathbf{F}$ is a vector of $n$ unknowns and $\mathbf{B}$ is a vector of $n$ values formed by the right-hand sides of the linear equations. In order to study the existence and uniqueness of the solution we consider a complementary problem, namely an eigenvalue problem for the matrix $\mathbf{A}$,

$$(\mathbf{A} - \lambda \mathbf{I})\mathbf{x} = 0, \qquad (28)$$

where $\mathbf{I}$ is an $n \times n$ identity matrix, $\mathbf{x}$ is an eigenvector that corresponds to the eigenvalue $\lambda$. For the Helmholtz equation with the same boundary conditions and similar finite difference discretization such analysis has been performed in [71] using the Matlab's routine *eigs*. For our investigation, we have developed a computer code for numerical solution of (27), where the eigenvalues and eigenvectors of the matrix $\mathbf{A}$ are calculated by calling the subroutine CGEEV of the LAPACK (Linear Algebra Package) [72]. As an example, in Fig. 4 and Fig.5 we show some eigenvalues and eigenvectors of the matrix which corresponds to the illustrative problem used throughout this paper. It is important to note that $\lambda = 0$ is not among the eigenvalues of the matrix. Thus, for $\lambda = 0$ the equation (28) has only a trivial solution for $\mathbf{x}$. Therefore, according to the fundamental theorems of linear algebra [73] (see [71] for detailed references to them) the equation (27) has a unique solution for any nonzero $\mathbf{B}$.

For solution of the equation (27) the code uses the subroutines BANDEC and BANBKS (LU decomposition, forward– and back–substitution of band diagonal matrices) [74], modified for complex matrices. For cross–checking of the results the code includes as an option the possibility to solve the linear system of equations using the LAPAC package (subroutine CGESV). The carried out benchmark numerical experiments have demonstrated that the results provided by both algorithms are practically identical. It should be mentioned, however, that the first one is more concise and faster, which makes it more suitable for embedding in other cavity codes belonging to GYROSIM package. Different implementations of both the model and the code have been compared as well. For example, a series of calculations have been performed in order to study the influence of the discretization of the boundary conditions on the solution. They have revealed that if one–sided differences are used instead of the ghost point method the field profile exhibits significant fluctuations close to the output section of the cavity, which we consider as unphysical. Another important parameter that influences the accuracy of the solution is the step of the discretization $\Delta z$. A care has been taken to ensure that the number of points per wavelength ($PPW = 2\pi / k_z \Delta z$) is sufficiently large. In typical numerical experiments the number of steps along the $z$ axis has been in the order of 800–1000.

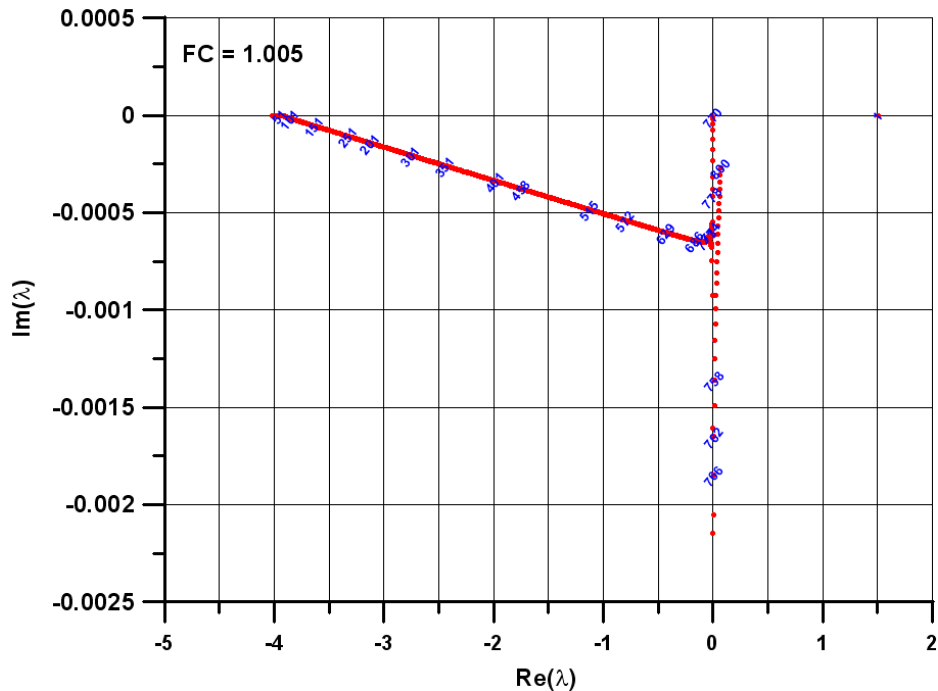

Fig. 4 First 800 eigenvalues of the matrix **A** calculated for a frequency $f / f_{cut} = 1.005$. The labels show the numbers of some of the eigenvalues.

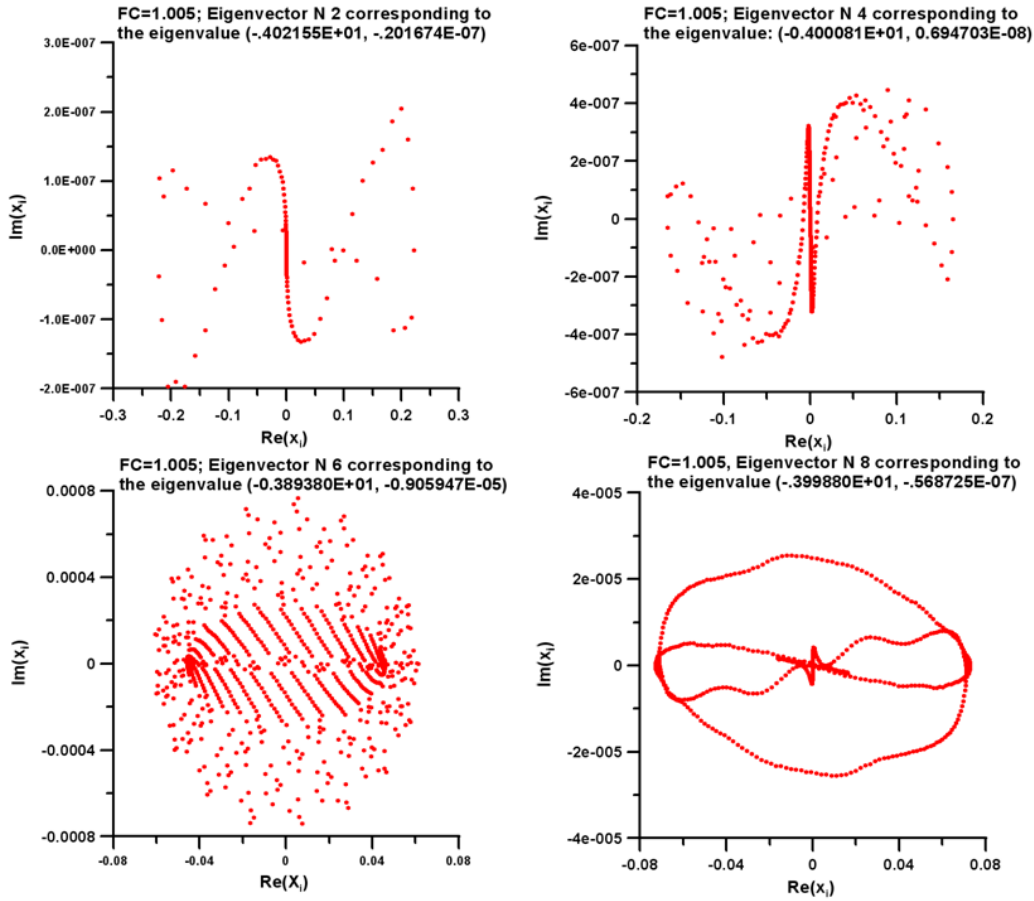

Fig. 5 Some eigenvectors corresponding to of the second, forth, sixth, and eight eigenvalue of the matrix **A**, respectively.

A series of axial field profiles normalized to the maximum value of their amplitudes and calculated for different frequencies are shown in Fig. 6. In these computations we consider the simplest case of a very small but different from zero excitation (since, as mentioned above, the homogeneous equation has only a trivial solution), which is constant inside the regular section of the cavity. In fact, for such formulation, the value of this constant affects only the magnitude of the profile but not its structure. The sequence in Fig. 6 shows how the standing wave (that is being formed as a result of the superposition of a forward and a backward propagating wave in the cavity) changes with frequency. Different profiles can be distinguished by the number of peaks, their relative amplitudes, shape, position, and spatial extend.

The bottom line, of this section, is that the finite difference discretization (27) of the physical model, which includes the inhomogeneous Helmholtz equation (25) supplied with the boundary conditions (6), provides solutions for the field profile at any real frequency. Therefore it can be used to analyze the spectral response of the cavity for a continuous spectrum of frequencies and to identify the resonances of HOAM.

## 5. Finite-bandwidth resonances of high order axial modes in a gyrotron cavity

In this section we discuss several resonance curves (spectral lines) that can be used to identify finite-bandwidth resonances of high–order axial modes in an open gyrotron cavity. Our approach here is similar to the spectral domain analysis and the field energy method proposed in [75–77].

Fig. 7 presents the first of the considered resonance curves, namely the dependence of the maximum field amplitude of the axial field profiles (some of which were shown in Fig. 6) on the frequency. Here and in the next figures the frequency $f$ is normalized to the cut-off frequency $f_{cut}$ of the cavity and all curves are normalized to the corresponding maximum value in the considered frequency range. The extrema (peaks and troughs) are indicated by vertical lines. Some of the peaks are marked by Gaussian profiles (green lines) plotted from three–point fits using the adjacent points to the maximum. Due to the overlapping of the profiles such an approximation is not completely adequate. We use it here only to illustrate graphically that the higher the axial mode the wider is its resonance curve. It is interesting to point out that this observation is consistent qualitatively with the results for the resonance curve in an open tube, obtained in [78] from a rather different point of view, namely from the diffraction theory at an open end of a waveguide. In this pioneering paper it is shown that for an open tube the width of the $q$-th resonance is proportional to $q^2$ while the distance between the adjacent peaks is proportional to $2q \pm 1$ (see Fig.5 in [78]).

Plots of several other quantities that depend strongly on the frequency and exhibit clear maxima can also be used as resonance curves. Some of them are shown in Fig. 8. The red curve is the squared field amplitude $F_{max}^2$ normalized to the maximum value. The integrals $\int_{z_{in}}^{z_{out}} \hat{F}^2(z)dz$ of the squared amplitude $\hat{F}^2(z)$ of the normalized field profiles $\hat{F}(z)$ are represented by the black line. This quantity is proportional to the storied energy $W$ divided by the factor $F_{max}^2$ since, by definition,

$$W = \frac{\varepsilon_0}{2} F_{max}^2 \int_{z_{in}}^{z_{out}} \hat{F}^2(z)dz, \tag{29}$$

where $\varepsilon_0$ is the permittivity of vacuum. The quantity $QP_{out} = \omega W$ ($Q$ and $P_{out}$ being the quality factor and the output power, respectively) is represented by the blue line. An enlarged part of these curves around the resonance of the second axial mode is shown in Fig.8b, where, additionally, a line which corresponds to the integral of the normalized field profile ( $\int_{z_{in}}^{z_{out}} \hat{F}(z)dz$ ) is plotted as well.

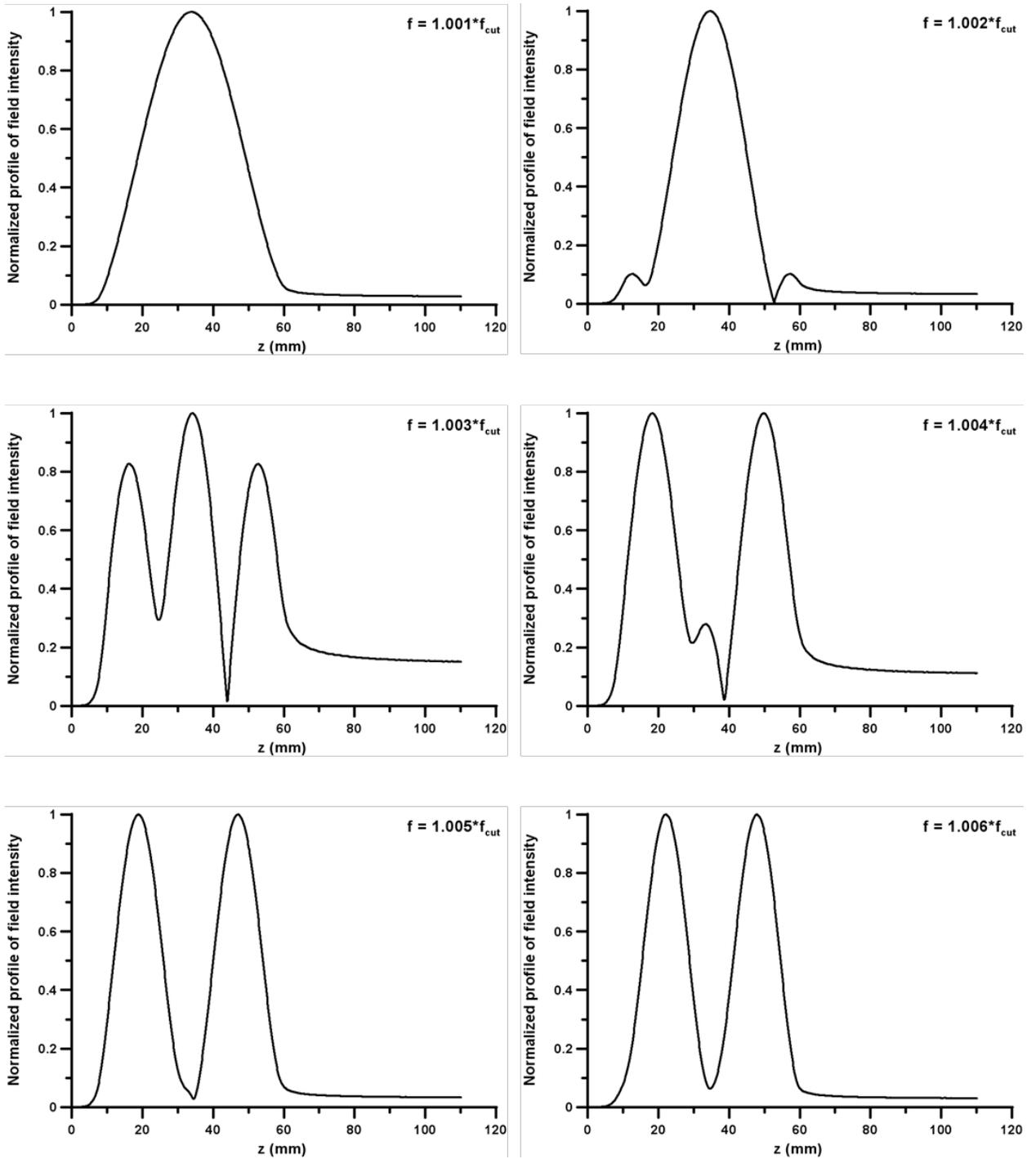

Fig.6 a) Normalized field profiles at frequencies in the range (1.001÷1.006).$f_{cut}$.

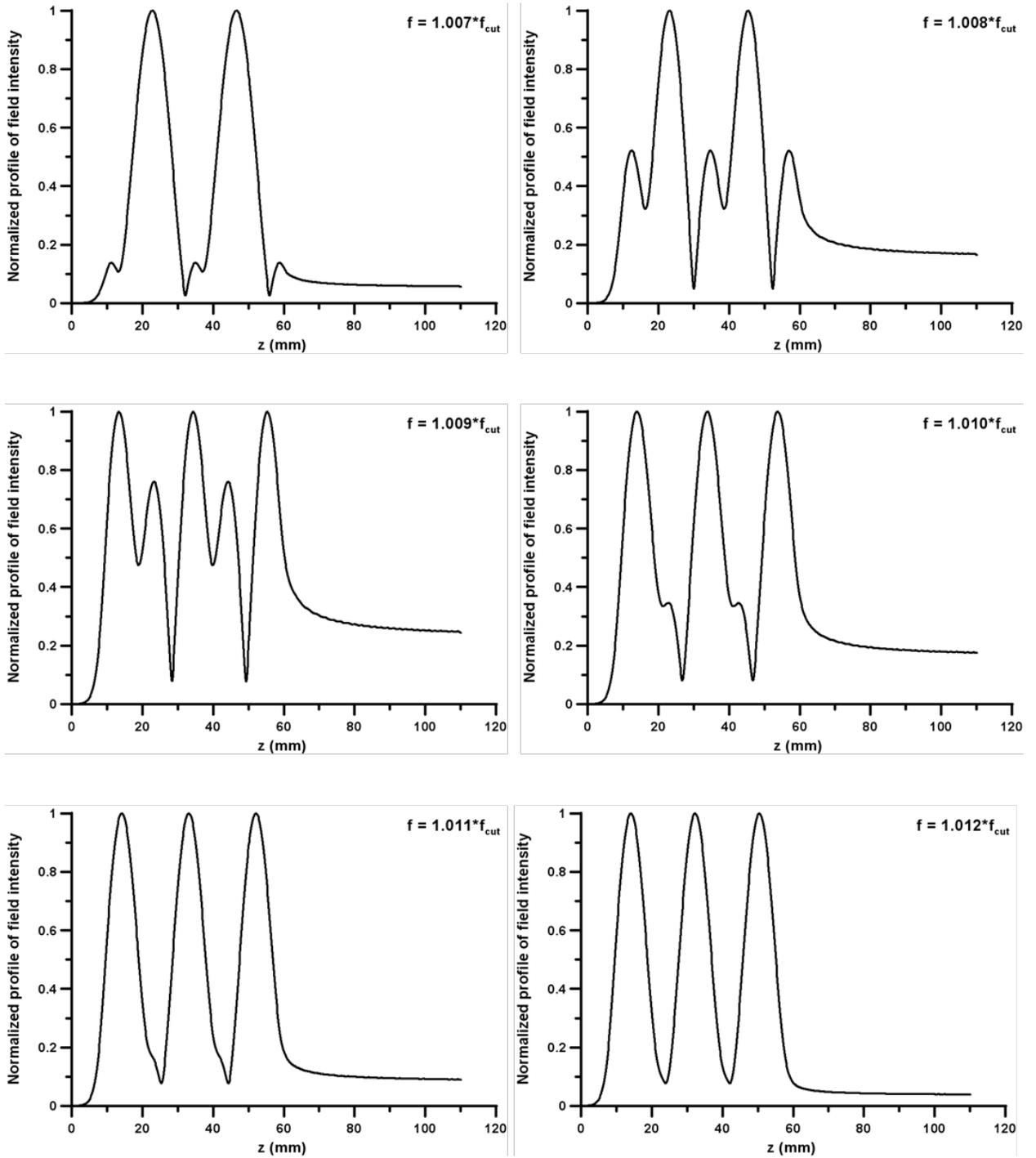

Fig.6 b) Normalized field profiles at frequencies in the range $(1.007 \div 1.012) \cdot f_{cut}$.

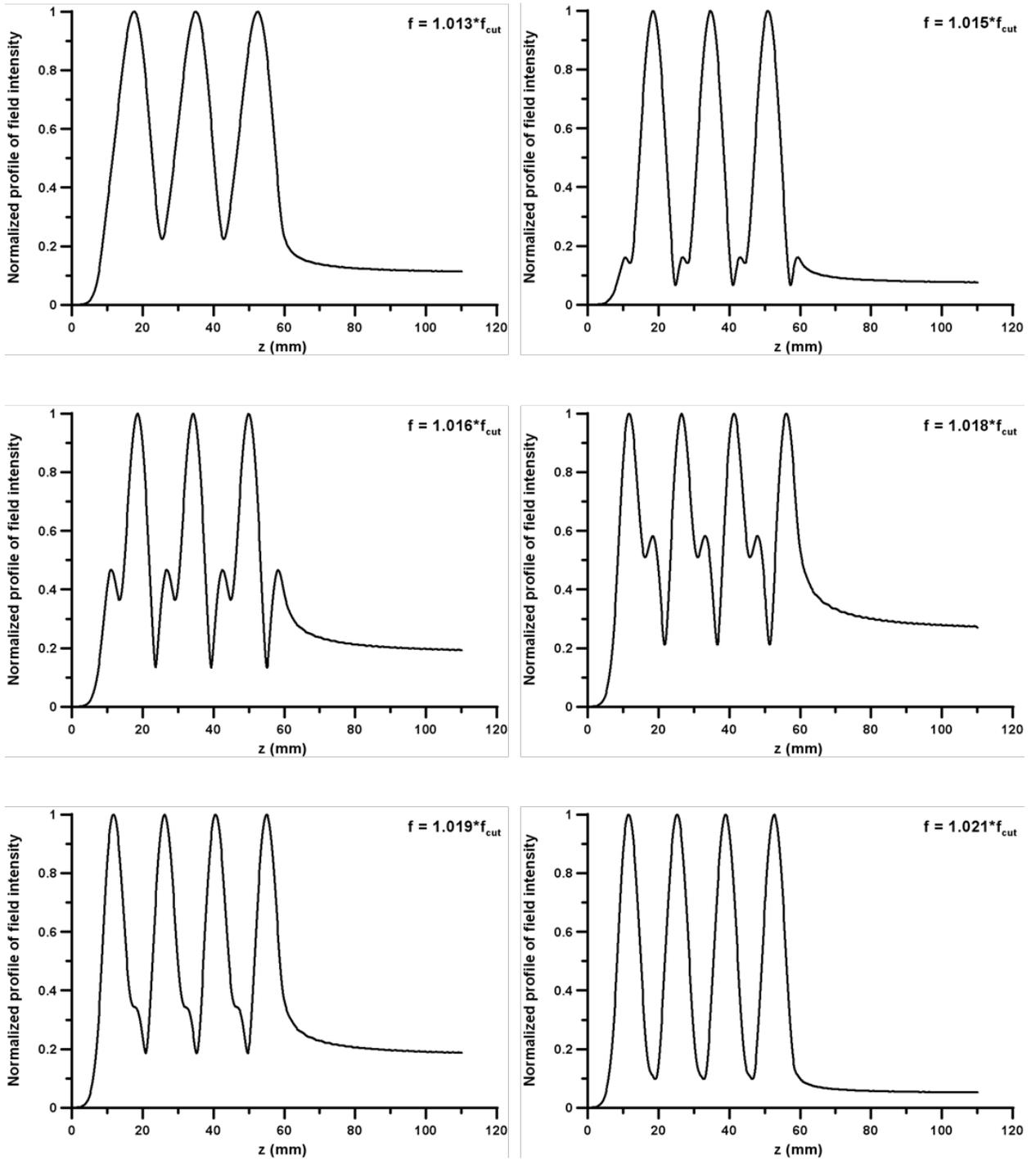

Fig.6 c) Normalized field profiles at frequencies in the range $(1.013 \div 1.021) \cdot f_{cut}$.

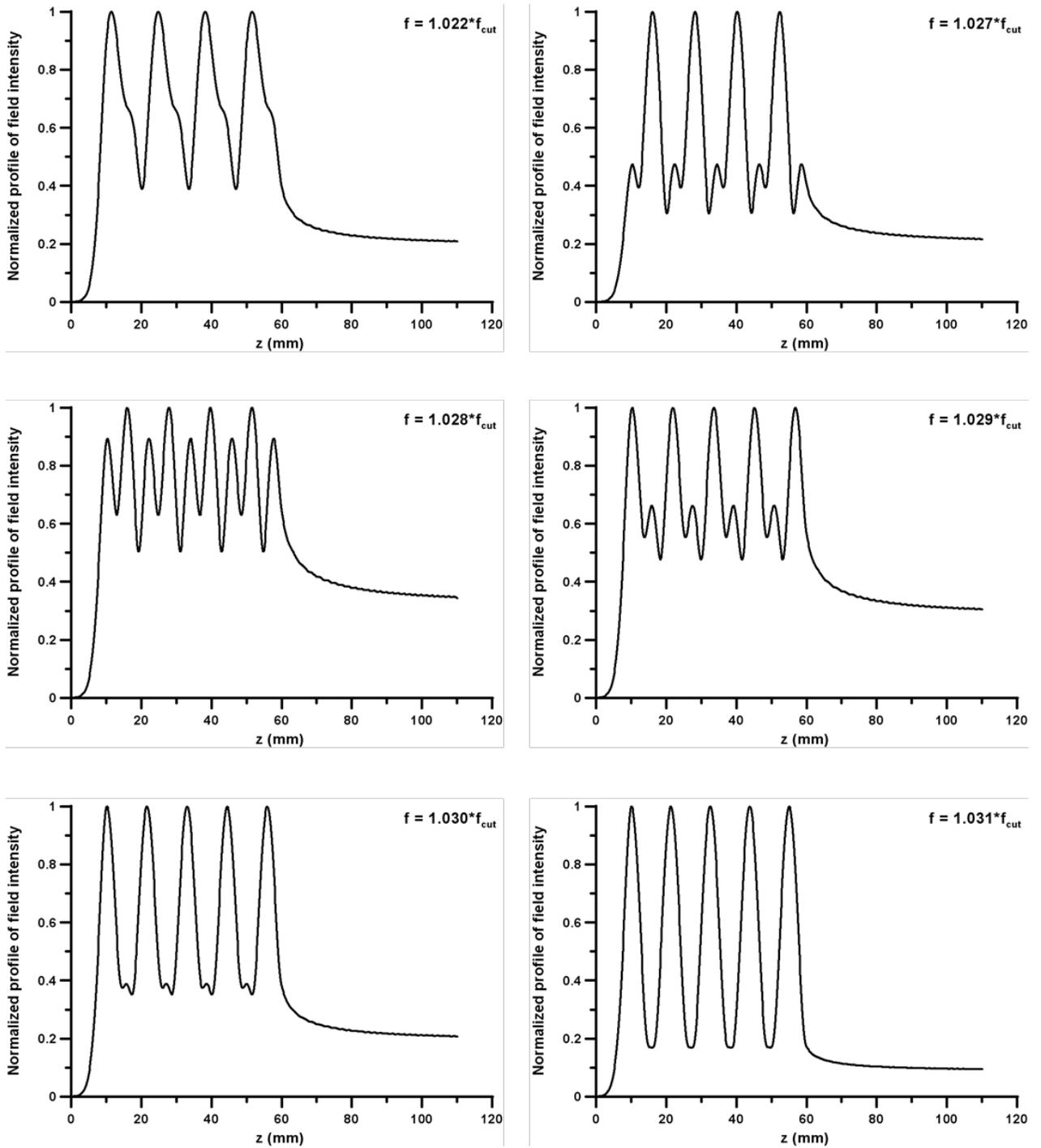

Fig.6 d) Normalized field profiles at frequencies in the range $(1.022 \div 1.031) \cdot f_{cut}$

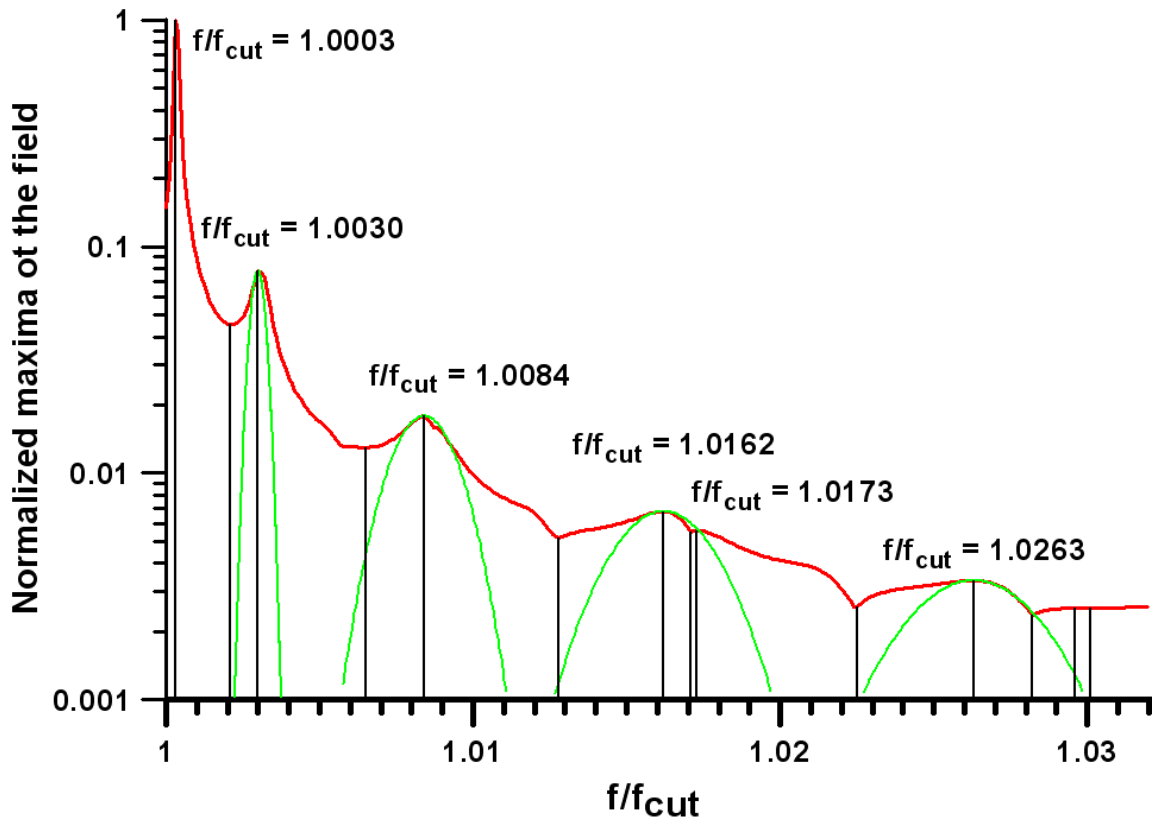

Fig. 7 Normalized maxima of the field profiles as a function of the frequency (red line). The vertical lines indicate the position of the extrema (peaks and troughs). Gaussian fits of the resonance peaks are plotted as green lines.

Analogous resonance curves, but plotted as functions of the axial wave number instead of the frequency are presented in Fig.9. Although their connotation is the same, the shapes of the peaks are slightly different due to the nonlinear relation between the wave number and the frequency. It appears, that in some cases, the latter curves are more appropriate since their peaks are more distinct (better resolved).

The resonances identified from both the resonance curves (obtained using the direct solution of the BVP as described in Sec.4) and by the cold cavity codes (using the shooting method) are presented in Tab.1 and on the dispersion curve of the studied mode in Fig.10.

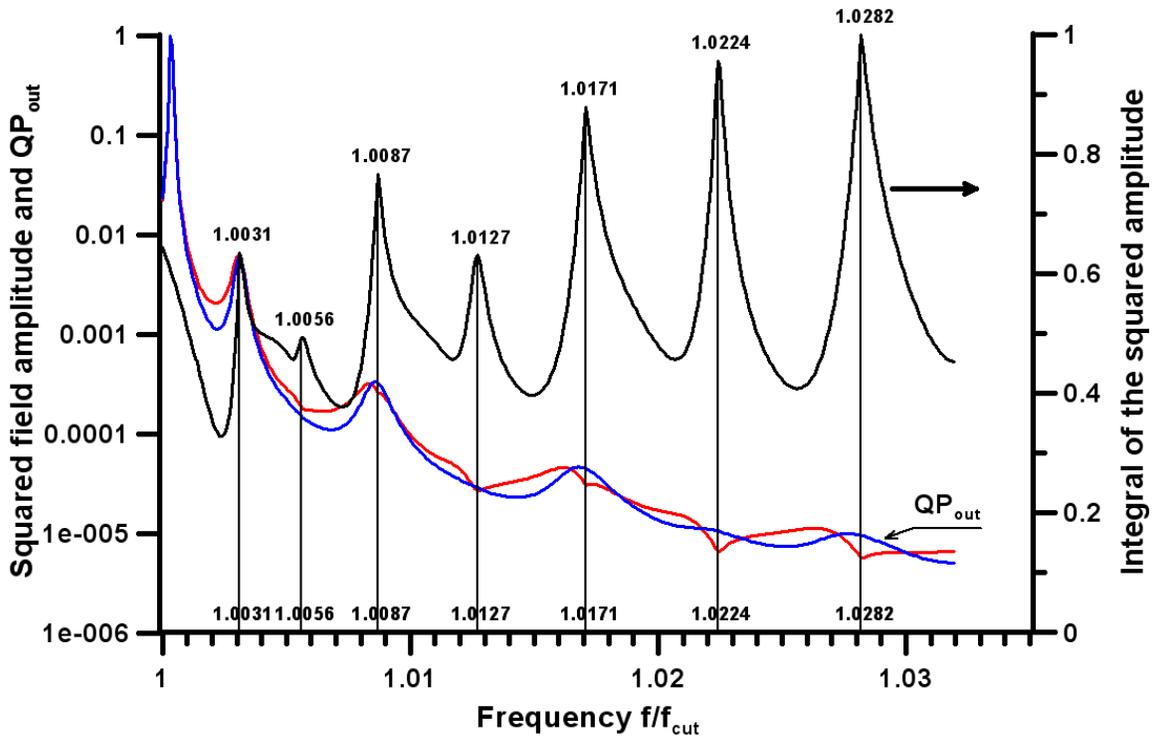

Fig. 8 a) Normalized resonance curves as functions of frequency: squared field amplitude (red line), $QP_{out}$ (blue line), and integral of the squared amplitude (black line). Vertical lines indicate the positions of resonances.

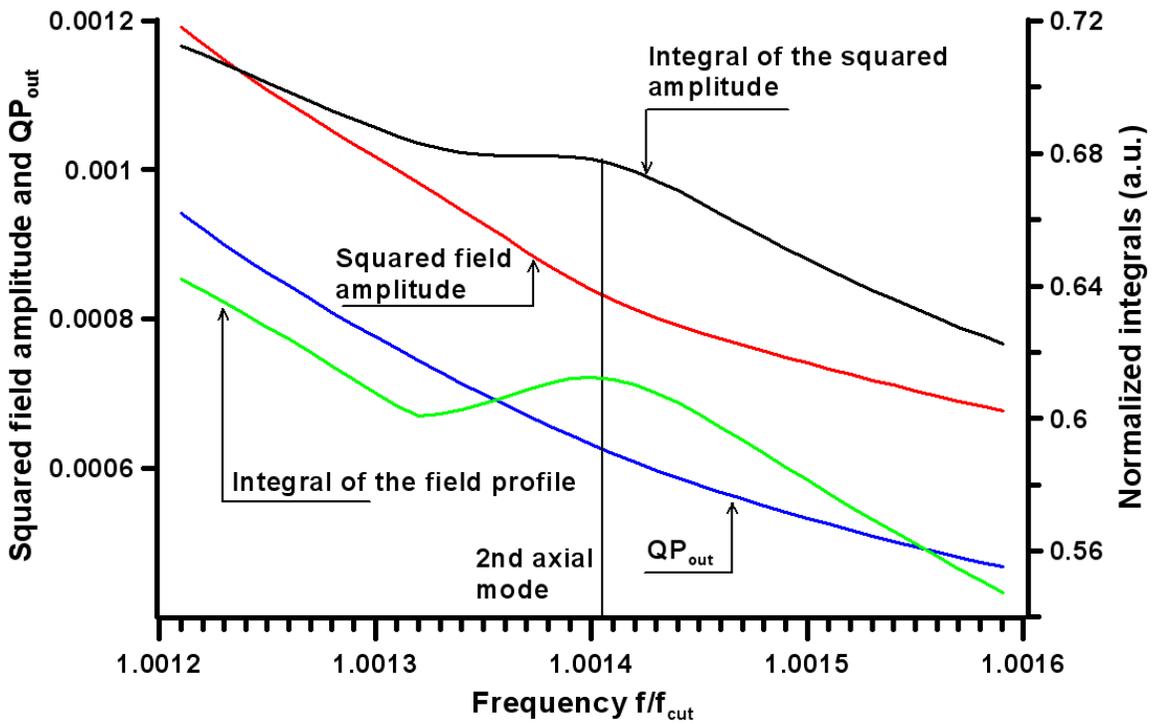

Fig. 8 b) Resonance curves of Fig. 8a) enlarged around the resonance of the second axial mode.

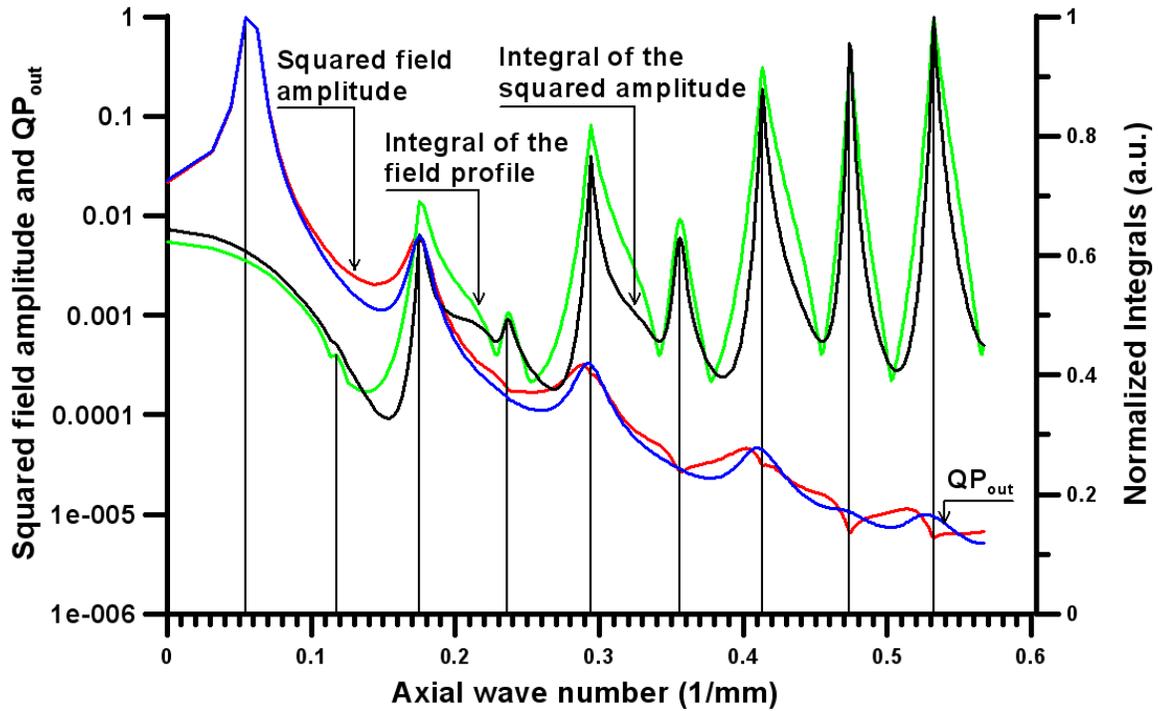

Fig. 9 a) Normalized resonance curves as functions of axial wave number: squared field amplitude (red line), $QP_{out}$ (blue line), integral of the squared amplitude (black line), integral of the field profile (green line). Vertical lines indicate the positions of resonances.

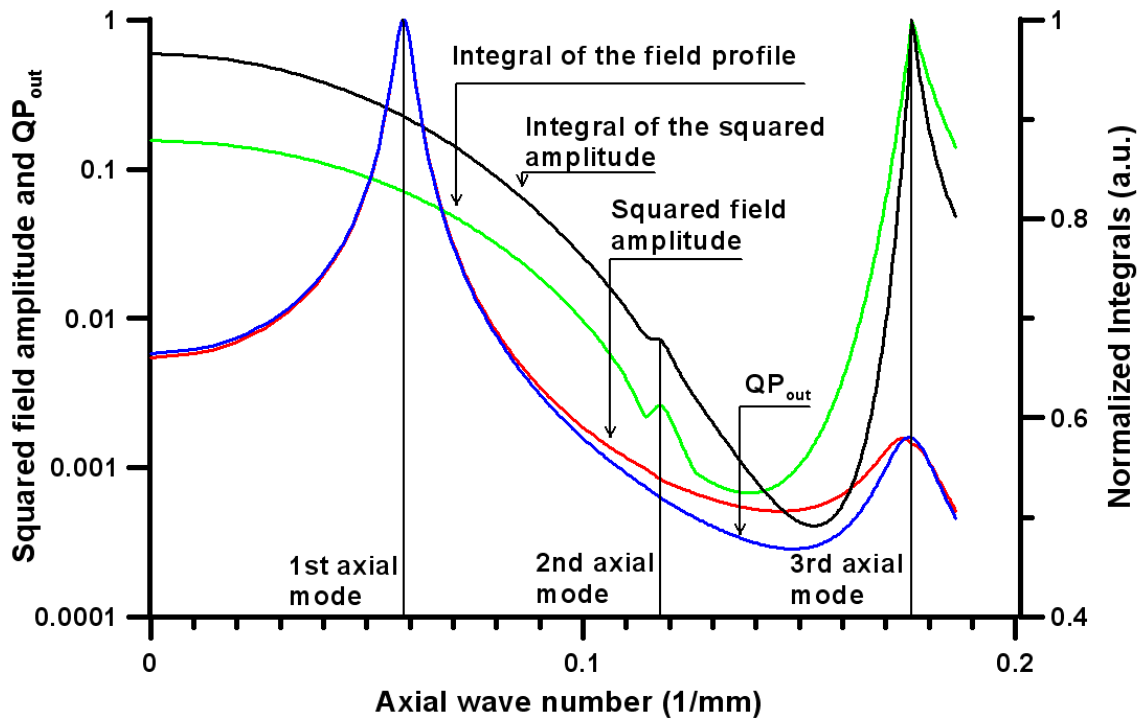

Fig.9 b) Resonance curves of Fig. 9a) enlarged around the resonance of the second axial mode.

Table 1 Comparison of resonances identified by different methods

| Resonances found by the cold cavity code (shooting method) | | | Resonances identified from the resonance curves (direct solution of the BVP) | | |
| --- | --- | --- | --- | --- | --- |
| $k_z$ (mm$^{-1}$) | f, GHz | f/f$_{cut}$ | $k_z$ (mm$^{-1}$) | f, GHz | f/f$_{cut}$ |
| 0.05846 | 106.1951 | 1.00031 | 0.0545 | 106.1904 | 1.00027 |
| 0.11692 | 106.3050 | 1.00135 | 0.1178 | 106.3103 | 1.00140 |
| 0.17538 | 106.4877 | 1.00307 | 0.1720 | 106.4802 | 1.00300 |
| 0.23383 | 106.7431 | 1.00548 | 0.2379 | 106.7604 | 1.00564 |
| 0.29224 | 107.0704 | 1.00856 | 0.2890 | 107.0503 | 1.00837 |
| 0.35060 | 107.4685 | 1.01231 | 0.3557 | 107.5100 | 1.01270 |

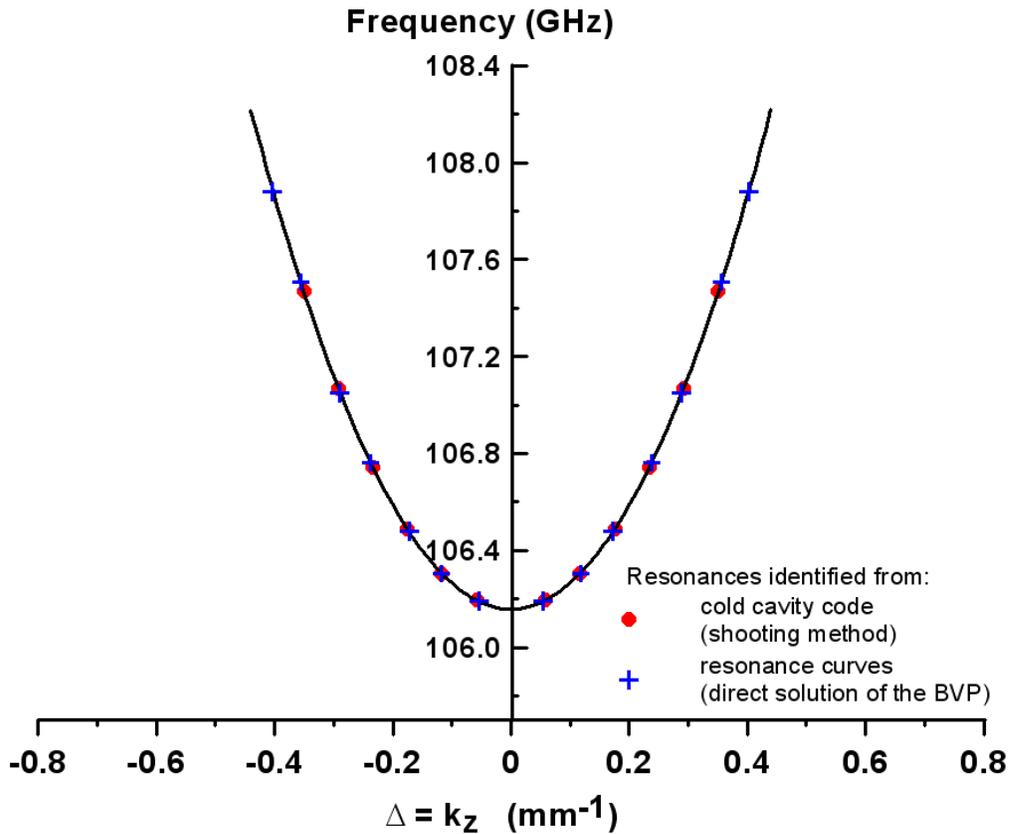

Fig. 10 Dispersion curve of the mode and the resonances identified from the cold cavity code (shooting method) – red dots, and from the resonance curves – blue crosses.

Their comparison shows that they are in close agreement. An important observation is that the first of the considered resonance curves (Fig. 7) allows one to identify only the odd resonances (1st, 3rd, 5th ... points in Fig. 3), while the resonance curves in Fig.8–9 indicate all resonances, including the 2nd, which (in this particular case) exhibits the weakest but still noticeable peak. The field profiles that correspond to the identified resonances are shown in Fig. 11. Comparing them with Fig. 2 it can be seen that the peaks corresponding to the odd resonances have equal number of peaks and similar shape while the profiles of the even resonances obtained from the cold cavity codes (shooting method) and by the direct solution of the BVP differ in this respect. We attribute the latter observation to the subtle differences between these two approaches for solving the BVP for the Helmholtz equation and more generally to the difference between the odd and the even resonances [78]. In this paper, however, we confine ourselves only to mentioning this peculiarity and intend to study and to comment it in more detail elsewhere.

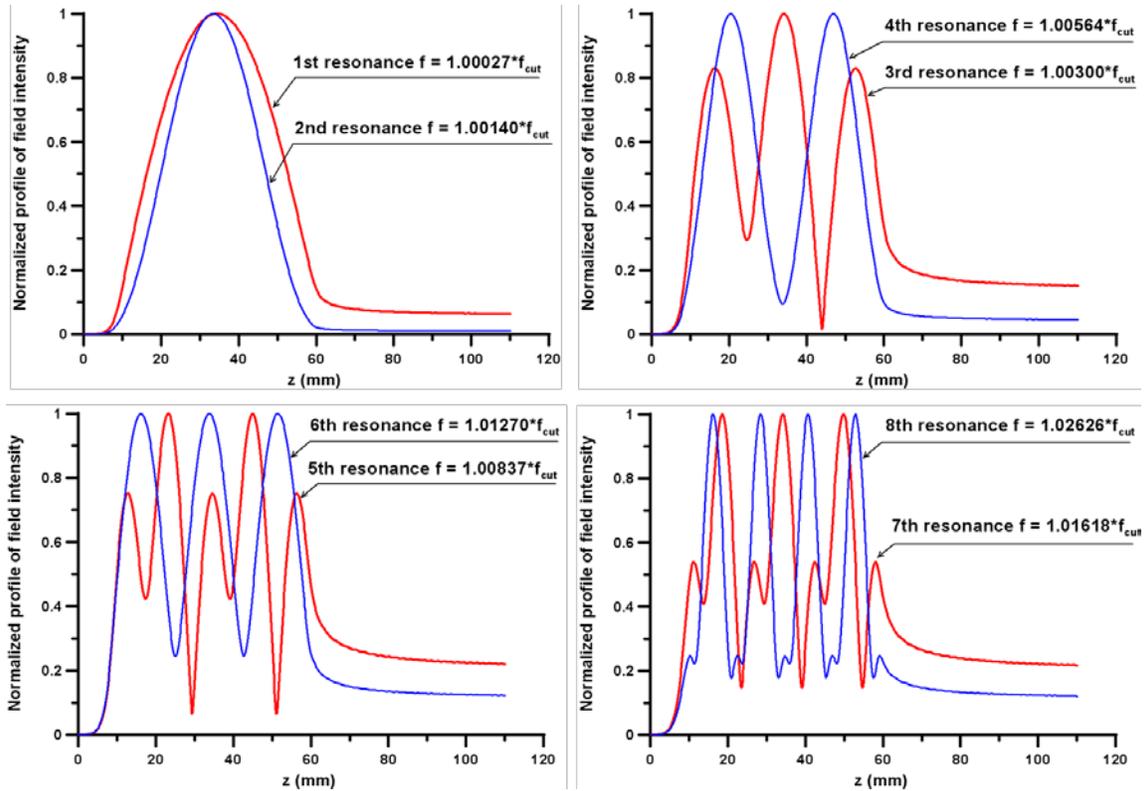

Fig. 11 Normalized profiles of the field intensity corresponding to the first eight resonances identified from the resonance curves in Fig. 7-9.

As it is evident from the spectral lines (i.e. curves in Fig. 8-9 that represent the frequency response of the resonator), the resonances of the HOAM in an open cavity have finite widths that widen as the ordering numbers of the resonances increase, which leads to their overlapping. Taking into account the finite widths of the resonances the latter should be represented on the dispersion diagram by bands rather than by points as in the case of a closed cavity. Schematically, this concept is illustrated in Fig. 12 and is applied to the studied cavity in Fig. 13, where the resonance bands of some of the HOAM are shown. Obviously, the widths of these bands depend on the convention used in order to estimate them. The one, which is the most common in similar problems, is the full width at the half maximum (FWHM). Due to the overlapping of the resonances such evaluation is not possible without resolving the peaks. In our code, the peaks are fitted by Gaussian profiles $A(f) = A_{max}(f_{res})\exp\left[-\frac{(f-f_{res})^2}{\sigma^2}\right]$ and the half width of the resonance is calculated from the relation $\Delta f = \sigma\sqrt{-\ln C}$, where $C = A(f_{edge})/A_{max}(f_{res})$ is the fraction of the maximum at which the resonance curve is truncated (at a frequency $f_{edge}$). Instead of FWHM (for which $C = 0.5$ and $\Delta f = 0.83\sigma$) in Fig. 13 the bands are plotted for $C = 0.95$, and correspondingly, $\Delta f = 0.23\sigma$ since for smaller values of $C$ the peaks overlap. Another appropriate approximation for the peaks would be a Lorentzian function as proposed in [76]. Such technique has been applied to the resonance curve shown in Fig. 14. The fits plotted there have been calculated by the Levenberg-Marquardt method using the QtiPlot software [79].

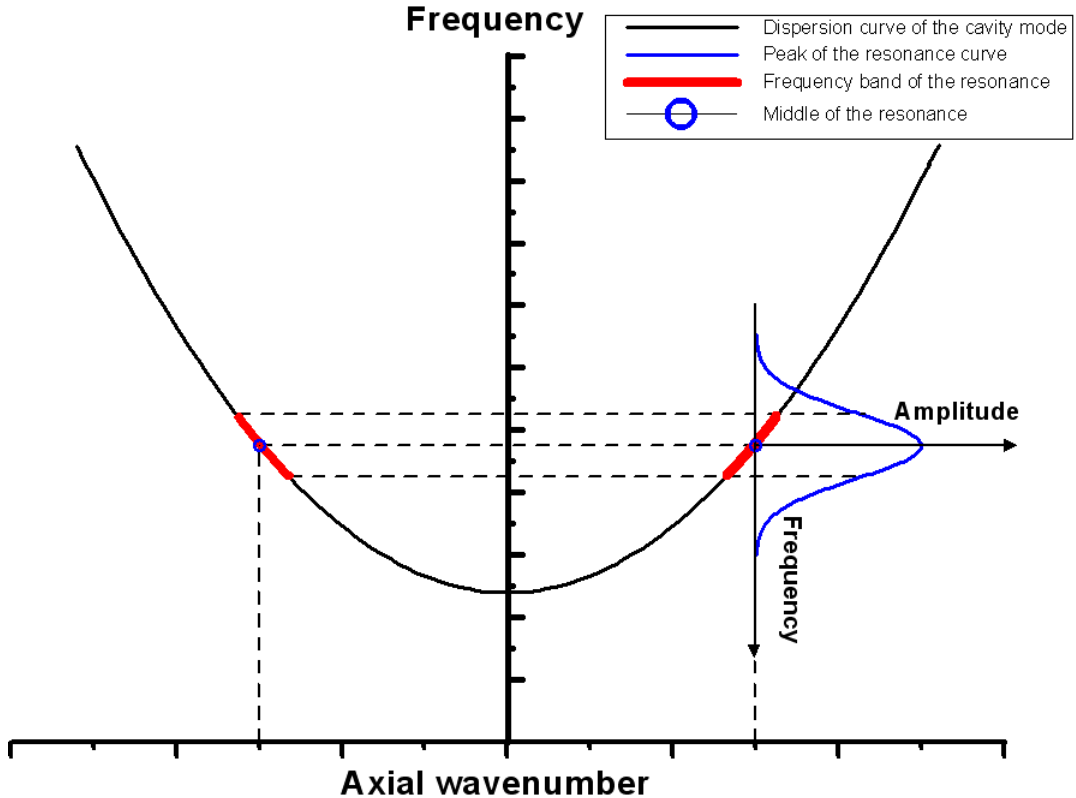

Fig. 12. Schematic representation of the concept of finite–bandwidth resonance bands.

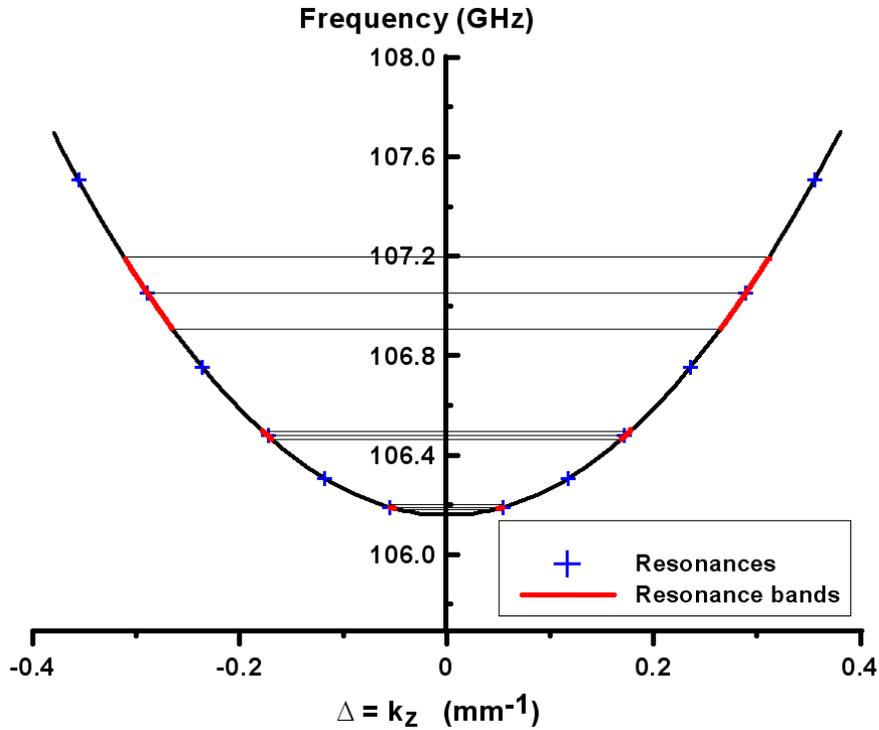

Fig. 13. Some finite−bandwidth resonance bands of the resonances shown in Fig. 10.

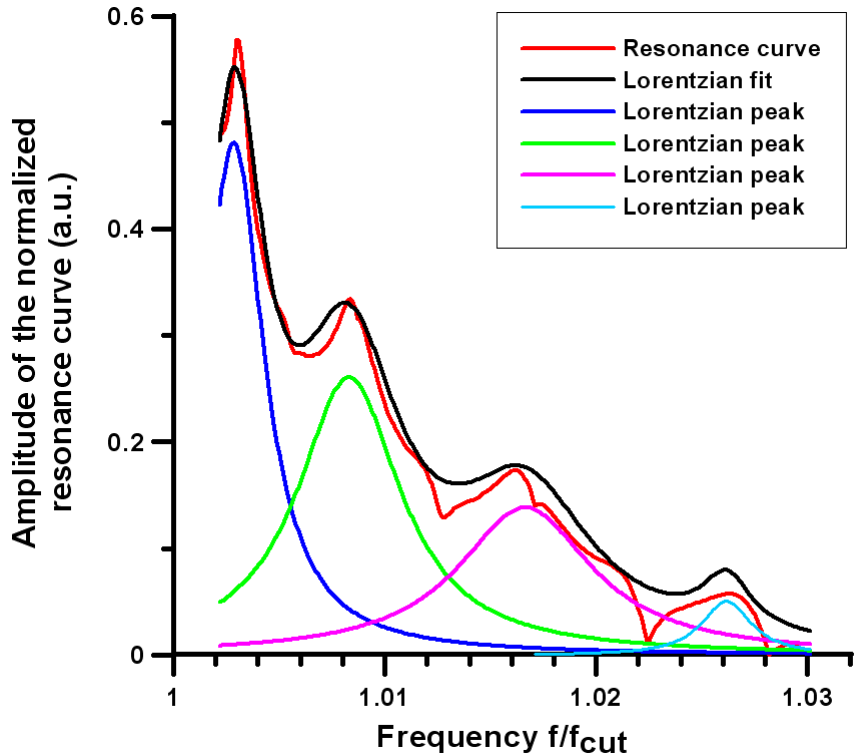

Fig. 14. Lorentzian fit of a part of the resonance curve (in this case, for convenience, the logarithm of the squared field amplitude, $\log_{10} F_{max}^2$ vs. frequency is used) normalized to unity at the maximum and the corresponding Lorentzian peaks.

The approach presented above can be used to study the influence of the geometrical parameters of the cavity on the spectrum of HOAM. As an illustration, in Fig. 15 and 16 we present the resonance curves of a modified cavity of Fig. 1, in which the length of the regular section is increased by 20 mm, other dimensions being the same. The resonance curves in Fig. 15 and Fig. 16 are plotted as functions of the frequency and the wave number, respectively. Additionally, in Fig. 16 the upper abscissa indicates the axial indices calculated as described above. It has already been mentioned that they should be considered as effective axial indices ($q_{eff}$). They are intentionally shown here to illustrate the fact that their values at the peaks of the resonances are not always integer numbers. The resonances identified for this longer cavity are presented in Tab. 2.

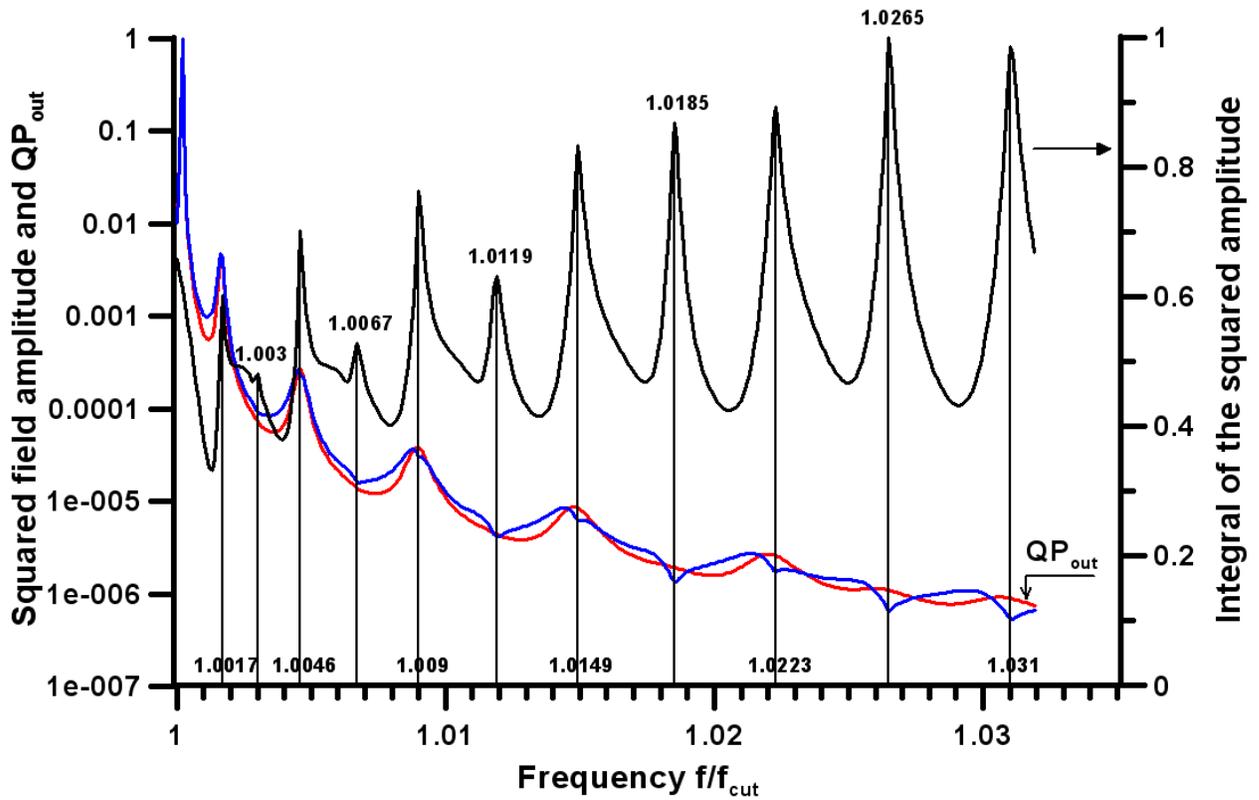

Fig. 15. Normalized resonance curves as functions of frequency for the modified (longer) cavity: squared field amplitude (blue line), $QP_{out}$ (red line), and integral of the squared amplitude (black line). Vertical lines indicate the positions of resonances.

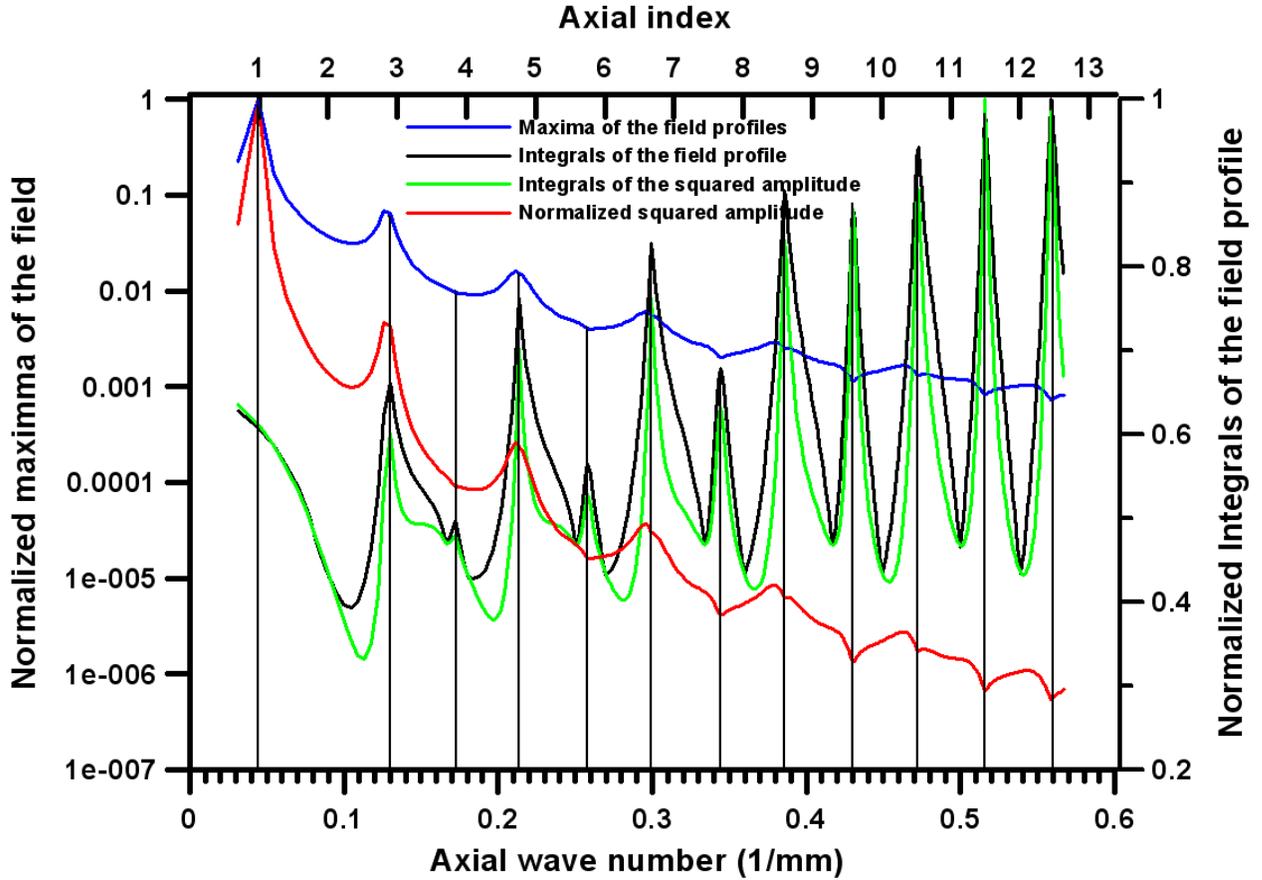

Fig. 16 Resonance curves as functions of the axial wave number for the modified (longer) cavity.

Table 2 Comparison of resonances identified by different methods in a longer cavity

| Resonances found by the cold cavity code (shooting method) | | | Resonances identified from the resonance curves (direct solution of the BVP) | | |
|---|---|---|---|---|---|
| $k_z$ (mm$^{-1}$) | f, GHz | f/f$_{cut}$ | $k_z$ (mm$^{-1}$) | f, GHz | f/f$_{cut}$ |
| 0.04262 | 106.1780 | 1.00015 | 0.04450 | 106.1797 | 1.00020 |
| 0.08525 | 106.2364 | 1.00070 | 0.12980 | 106.3390 | 1.00170 |
| 0.17049 | 106.4697 | 1.00290 | 0.17249 | 106.4770 | 1.00300 |
| 0.21310 | 106.6443 | 1.00455 | 0.21367 | 106.6469 | 1.00460 |
| 0.25571 | 106.8573 | 1.00655 | 0.25801 | 106.8699 | 1.00670 |
| 0.29830 | 107.1084 | 1.00892 | 0.29920 | 107.1141 | 1.00900 |
| 0.34087 | 107.3972 | 1.01164 | 0.34429 | 107.4220 | 1.01190 |

In all examples presented and discussed above the excitation of the cavity (RHS of Eq.25) was a constant inside the regular section of the cavity. The value of this constant influences only the amplitude of the field profiles, while the shape of the normalized profiles and resonance curves (as well as the corresponding resonances) remain the same. It should be made clear that this constant could be made arbitrarily small (in which case we have practically a cold cavity since the excitation is negligible) but different from zero, since as it has been underlined in Sec. 4., the homogeneous equation has only a trivial solution. This fact was corroborated by the numerical calculations with different values of the constant excitation ranging from $10^{-12}$ to $10^{12}$.

Next, we present results for the case when the excitation of the cavity is a Gaussian function as shown in Fig. 17. The corresponding resonance curves are plotted in Fig. 18. For comparison, the lines obtained for a constant excitation are also reproduced there. With the same aim in view, in Fig. 19 we show several field profiles calculated at both a Gaussian and a constant (uniform) excitation. It can be seen that, as expected, the change in the excitation alters both the spectrum of the cavity and the corresponding field profiles.

Since the aim of this paper is to present and to demonstrate the developed approach, rather than to apply it to a detailed study of any particular case we confine ourselves only to these illustrative examples. It is clear, however, that it can be used in order to calculate the field profile at any excitation. We intend to embed the developed numerical modules in the cavity codes of our problem−oriented software package GYROSIM. The results of numerical experiments carried out with the upgraded computer programs will be published elsewhere in a sequel to this paper.

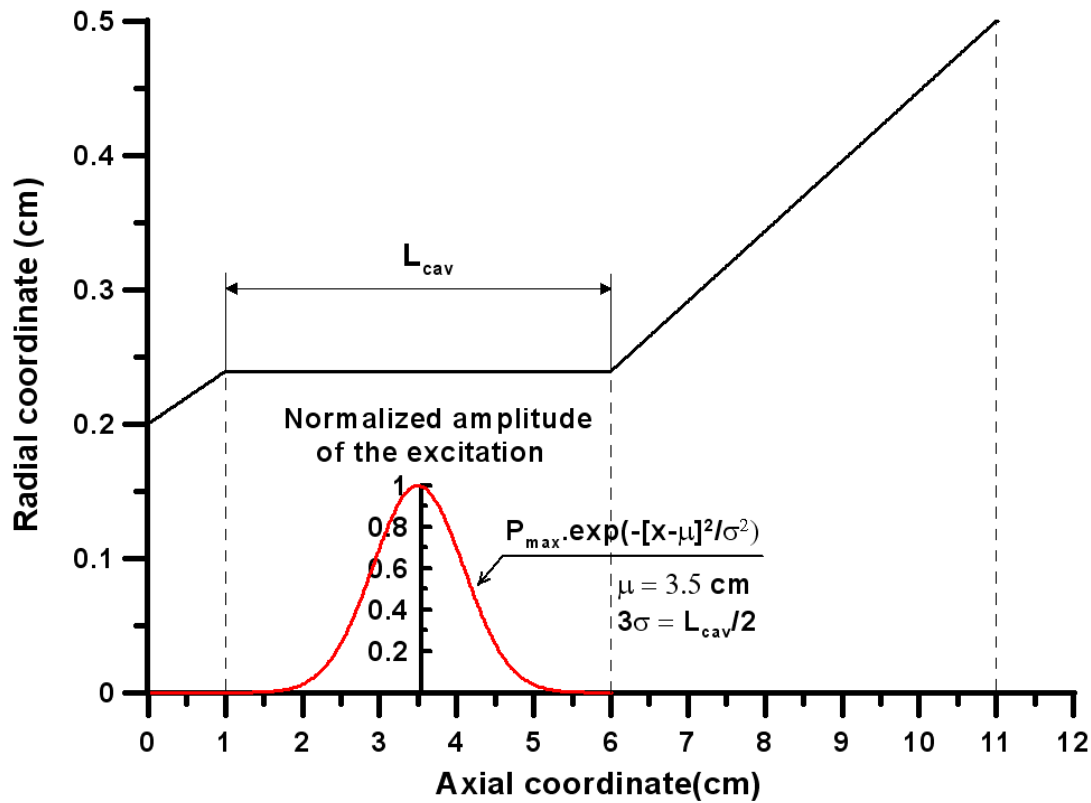

Fig. 17 Gaussian excitation of the resonant cavity

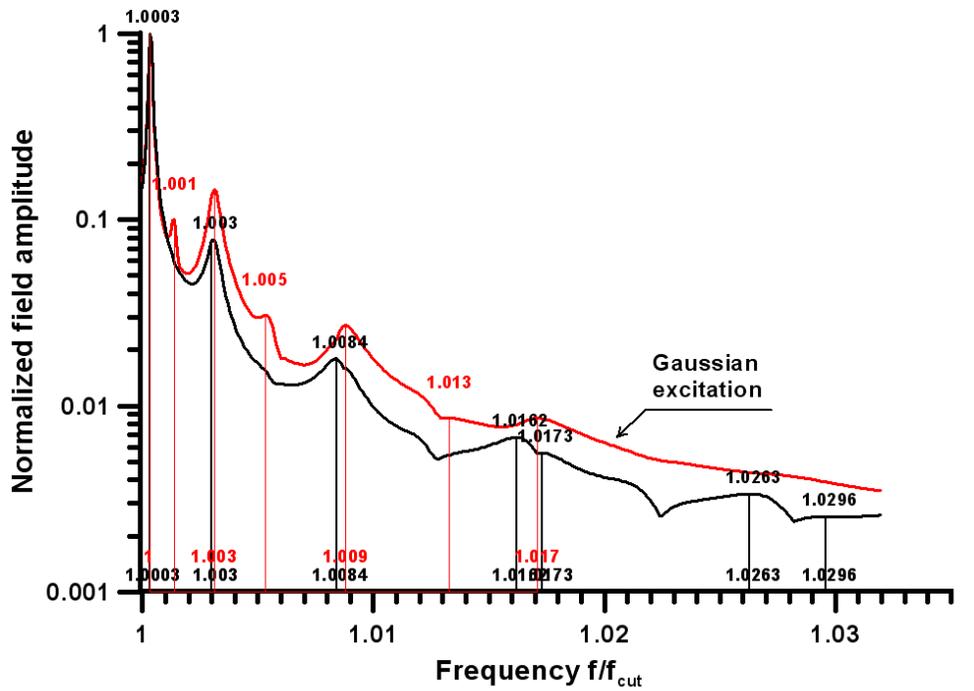

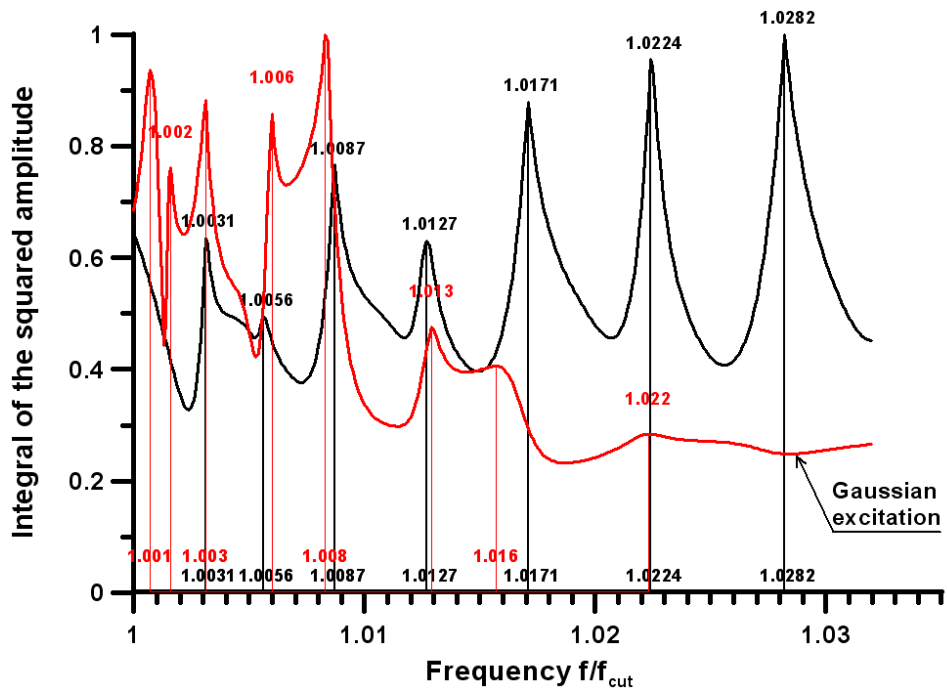

Fig 18 Comparison of the resonance curves for constant (black lines) and Gaussian (red lines) excitation of the cavity: a) normalized field amplitude vs frequency; b) integrals of the squared amplitude vs frequency.

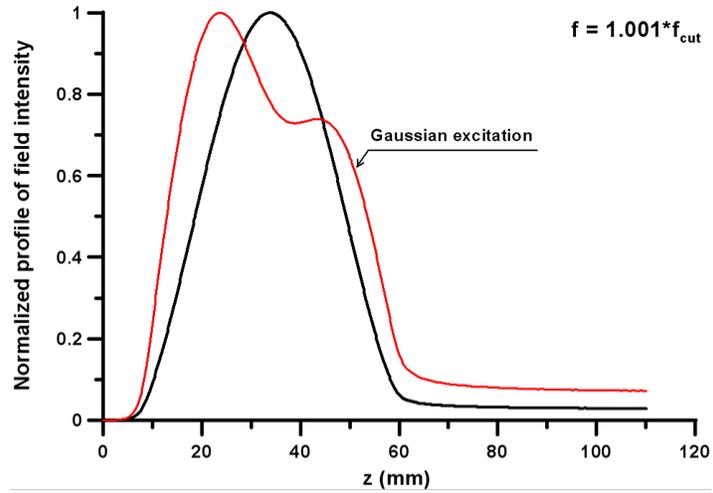
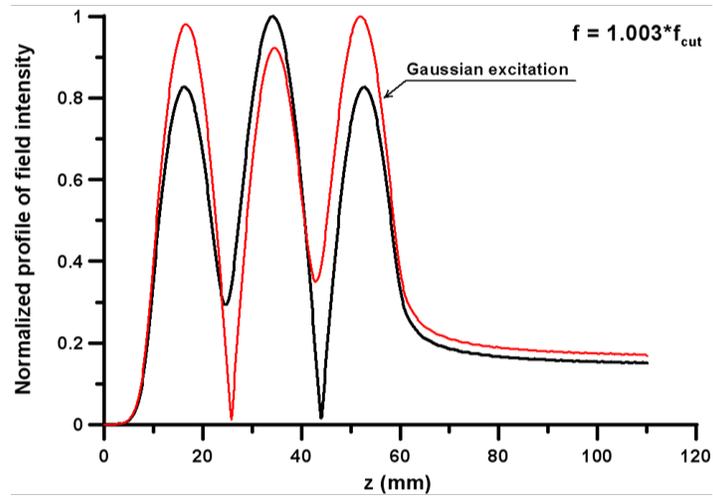
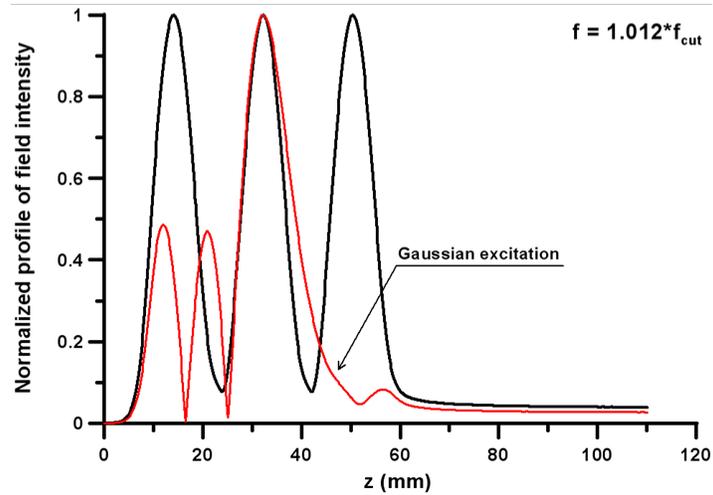

Fig. 19 Comparison between several field profiles calculated at constant (black lines) and Gaussian (red lines) excitation.

## 6. Conclusions

As in any resonator with losses, in a typical gyrotron cavity the resonances have finite widths (and, *a fortiori*, finite Q factors) due to both the energy escape from the open end and the Ohmic losses owing to the wall resistivity. In many studies, however, this well-known fact is very often neglected and the resonances are being represented by a discrete points (corresponding to the central frequency of the resonance) on the dispersion diagram of the cavity mode. In this paper a numerical approach has been presented which allows one to study the frequency response of the cavity (spectral analysis) and to identify finite–bandwidth resonances of HOAM. It is based on the usage of various resonance curves obtained from the solution of the boundary value problem for the inhomogeneous Helmholtz equation by the finite difference method (FEM). The results of the spectral analysis are compared with those from the eigen–solution of the homogeneous Helmholtz equation by the shooting method (Vlasov's approach). The developed computational modules will be integrated into the existing cavity codes of the problem–oriented software package GYREOSS in order to study the beam–wave interaction and energy transfer in gyrotrons operating on a sequence of (possibly overlapping) HOAM. The proposed approach can be extended further to more general boundary conditions (see Eq. (9)) and work on this is in progress now. We intend to publish the results of the numerical experiments, obtained by the novel (upgraded) codes elsewhere in a sequel to this paper.

## Acknowledgements

This work was supported partially by the Special Fund for Education and Research from the Ministry of Education, Culture, Sports, Science and Technology (MEXT) in Japan and was carried out in the framework of the collaboration of an international consortium on the project "Promoting international collaboration for development and application of high–power THz gyrotrons" of the FIR UF Research Center (Fukui, Japan).